\documentclass[nofootinbib,
	aps,
	prd,
	twocolumn]{revtex4-2}
\usepackage{mathtools}
\usepackage{mathrsfs}
\usepackage{amsfonts}
\usepackage{leftindex}
\usepackage{hyperref}
\usepackage{booktabs}
\usepackage{graphicx}
\usepackage[dvipsnames]{xcolor}
\usepackage{orcidlink}

\hypersetup{colorlinks=true,
			citecolor=blue,
			linkcolor=blue,
			urlcolor=blue!25!black
}
\newcommand{\dif}{\ensuremath{\mathrm{d}}}
\newcommand{\RR}{\mathbb{R}}

\newcommand{\GB}{\mathcal{G}}

\newcommand{\tr}{\text{tr}}
\newcommand{\real}{\text{Re}}
\newcommand{\imag}{\text{Im}}
\newcommand{\sgn}{\text{sgn}}

\newcommand{\cc}[1]{\overline{#1}}

\newcommand{\Hsigma}{\mathcal{H}_\sigma}
\newcommand{\Husigma}{\mathcal{H}'_\sigma}
\newcommand{\Hsigmahat}{\mathcal{H}_{\hat\sigma}}
\newcommand{\Husigmahat}{\mathcal{H}'_{\hat\sigma}}

\graphicspath{{Images/}}

\begin{document}
	
\title{Kundt gravitational waves coupled with a non-Noetherian conformal scalar field}
\author{Pedro A. S\'anchez\,\orcidlink{0009-0008-7437-0561}}
\email{pedro.sanchez.s@cinvestav.mx}
\affiliation{Departamento de F\'{\i}sica, Centro de Investigaci\'on y de Estudios Avanzados del I.P.N., Apdo. Postal: 14-740, 07000, Ciudad de M\'exico, M\'exico.}

\begin{abstract}
Type N spacetimes of the Kundt class are presented as solutions to Einstein's equations sourced by a real scalar field whose equation of motion is conformally invariant and that generalizes the standard conformal scalar field. The specific spacetimes considered model gravitational waves with uniform and totally geodesic wave fronts, propagating in a maximally symmetric background, and are characterized by the value of their constant scalar curvature and by the so-called wave profile function. All subclasses of such spacetimes are analyzed. It is shown that the scalar field solutions generically divide into two branches, of which one has a strikingly different behavior from that of the standard conformal case. The scalar field contributes to the equation satisfied by the wave profile function by adding new singular terms. Closed form and mode solutions for the wave profile function are found for different scenarios.
The resulting energy-momentum tensor has a null eigenvector, but is more general than the pure radiation type usually coupled with this kind of spacetimes.
\end{abstract}

\maketitle

\section{Introduction}

Among spacetimes modeling gravitational waves, those which are Petrov type N \cite{petrov1954klassifikacya,*petrov2000classification} and which belong to the Kundt class \cite{kundt1961plane} stand out for their simplicity and for their physical and geometrical richness. Firstly, vacuum type N spacetimes exhibit appealing wave properties: deviation of their timelike geodesics has a pure transverse character (cf. \S 2-2.10 of \cite{ehlers1962exact}), their curvature tensor approximates at null infinity that of an isolated system under generic conditions \cite{sachs1961gravitational}, and its curvature invariants of order zero vanish (cf. \S 92 of \cite{landau1975classical}). Secondly, when the principal null direction (PND) of an algebraically special spacetime has vanishing optical scalars ---thus defining a Kundt class spacetime--- it becomes an eigenvector of the Ricci tensor and, via Einstein's equations, also of the energy-momentum tensor of the matter content. Electromagnetic radiation, null dust or a generic pure radiation matter (cf. \S 4.3 of \cite{hawking1973the} or \S 5.2 of \cite{stephani2003exact}) are some examples of this type of energy-momentum tensor. Thus, spacetimes that are type N with their PND, say $k$, defining a Kundt class spacetime come out as reasonable simplified models of gravitational waves. In fact, they have been studied as such for over half a century.

The first to be studied \cite{brinkmann1923riemann,*brinkmann1925einstein} were the so-called plane-fronted waves with parallel rays, or pp waves for short, which are given locally by
\begin{equation} \label{eq:pp-waves_metric}
ds^2_{pp} = H du^2 - 2 du dv + 2 \left| d \zeta \right|^2,
\end{equation}
with $k = \partial_v$ and $H$ a real-valued function constant along each null geodesic (cf. also \S 2-5 of \cite{ehlers1962exact} and references cited therein). They are geometrically characterized by the fact that $k$ is covariantly constant, justifying the term parallel rays; additionally, the surfaces at constant $u$ and $v$, called wave surfaces, have vanishing Gaussian and extrinsic curvature, justifying the term plane-fronted. Furthermore, their principal null direction turns out to be a Killing vector field (KVF). Later on, Kundt \cite{kundt1961plane} found the most general type N vacuum solutions. They describe also plane-fronted waves, but $k$ is not necessarily covariantly constant, neither a Killing vector field. 

Generalizing the vacuum solutions, all type N Kundt class spacetimes with a cosmological constant $\Lambda$ and without additional matter are known. All of them feature uniform and totally geodesic wave surfaces, having constant Gaussian curvature, determined by $\Lambda$, and vanishing extrinsic curvature. They have the intriguing property that the third Lie derivative of the metric with respect to $k$ vanishes; this might happen if the Lie derivative with respect to $k$ acting twice on the metric results in zero or not. The latter case yields solutions for any $\Lambda$ \cite{garcia1981all}, featuring spherical, hyperbolical or plane wave fronts, while the former case admits solutions only for $\Lambda \leq 0$ \cite{ozsvath1985plane}, featuring only hyperbolical or plane wave fronts. The solutions not admitting spherical wave fronts contain as a particular case those for which $k$ is a KVF, which for $\Lambda < 0$ correspond to Siklos waves \cite{siklos1985lobatchevski} and are conformal to pp waves. Moreover, the solutions admitting spherical wave fronts are also conformal to their vacuum relatives.

All type N spacetimes of the Kundt class, exhibit a function $H'$ which is constant along each null geodesic, called wave profile function, analogous to the one appearing in \eqref{eq:pp-waves_metric}. For the $\Lambda$-vacuum solutions, it determines completely the Weyl tensor and is subject to
\begin{equation} \label{eq:wave-eq_Hp}
\Box H' = 0.
\end{equation}
Given the dependence of $H'$ and the structure of the metric, the foregoing wave equation yields an elliptic equation on the wave surfaces, reducing for pp waves to Laplace's equation on a plane. For a non-vanishing $\Lambda$ the general solution is given in terms of an arbitrary holomorphic function $F = F(u,\zeta)$ which cannot be eliminated through coordinate transformations. As noted in \cite{trautman1962propagation}, the appearance of an arbitrary function enables information propagation, a characteristic feature of waves. Remarkably, these $\Lambda$-vacuum solutions have been identified as those type N spacetimes for which any symmetric conserved tensor constructed from the metric, the curvature tensor and its covariant derivatives is a multiple of the metric itself, thus called \emph{universal} spacetimes \cite{hervik2014type}.

In addition to the solutions considered above, there exist other Kundt class spacetimes coupled to a wide variety of fields. General type N spacetimes of the Kundt class minimally coupled with a Maxwell field and with an Abelian Higgs field were studied in \cite{ozsvath1985plane} and \cite{canfora2021exact} respectively, and spacetimes in which $k$ is a KVF non-minimally coupled with a real scalar field and an electromagnetic field were found in \cite{ayon2007higher} and \cite{gurses1978pp} respectively. Broader spacetimes of the Kundt class have also been studied as solutions to alternative theories of gravity including those non-minimally coupled with an electromagnetic field, Lovelock theory and $f(R)$ theories \cite{horndeski1979null,ortaggio2018lovelock,svarc2020kundt,gurses2022kerr,baykal2022kundt}. The majority of the solutions that include matter have as a common feature an energy-momentum tensor of the pure radiation type. A more comprehensive list of investigations of Kundt class spacetimes can be found in \cite{baykal2022kundt}.

Nevertheless, despite the conformal properties of known type N Kundt solutions, there has not been so much study coupling Kundt spacetimes with matter having specific conformal properties other than the seminal works of \cite{ozsvath1985plane} and \cite{ayon2007higher}. Thus, we aim to study type N spacetimes of the Kundt class as solutions to Einstein's equations, sourced by the simplest possible matter content, namely a real scalar field, whose equation of motion is conformally invariant and for which all field equations are of second order. The most general action principle for such a field was recently found by Fernandes \cite{fernandes2021gravity} and includes as a particular case the previously known conformally invariant real scalar field (cf. \cite{gursey1963reformulation,penrose1964conformal, *penrose2011republication,callan1970new}). 
Since the resulting field equations are of second order, the action belongs to Horndeski's general class of scalar-tensor theories \cite{horndeski1974second}. If higher order terms are allowed in the energy-momentum tensor, the most general action principle for such a field was given in \cite{ayon2024non}.
This generalized conformal scalar field has been studied in the context of black holes, cosmology, Vaydia spacetimes and wormholes \cite{fernandes2021gravity,babichev2022conformally} for the second order case, while stealths on flat spacetime have been found for the higher order case \cite{ayon2024cheshire}. The present work will enhance the family of known solutions to the second order theory, to include those coupled with type N Kundt spacetimes.

This work is organized as follows. In \S \ref{sec:type-N_Kundt} we will describe type N spacetimes of the Kundt class, stating its defining and characteristic properties, presenting the closed form for the general class and reviewing the $\Lambda$-vacuum solutions. Most of the content of \S \ref{sec:type-N_Kundt} is a review, except for the general closed forms of the metric given for the general type N spacetimes of the Kundt class. Then, in \S \ref{sec:generalized-scalar-field} we will describe the generalized conformal scalar field that we will consider. Next, in \S \ref{sec:solutions} we will present the solutions found, beginning with a pp wave spacetime and later considering more general type N Kundt spacetimes. Finally, in \S \ref{sec:discussion} we will summarize and discuss the results obtained and comment on possible future work.

\section{Type N Kundt spacetimes} \label{sec:type-N_Kundt}

A type N spacetime of the Kundt class is characterized by a non-vanishing Weyl tensor and by the existence of a vector field $k$ satisfying
\begin{subequations} \label{eq:kundt-N_conditions}
\begin{align}
&k_a k^a = 0, &&k_{a;b} k^b = 0, &&k_{a;b} k^{a;b} = 0, \label{eq:kundt_def}\\
&C_{abcd} k^d = 0, \label{eq:type-N_def}
\end{align}
\end{subequations}
namely $k$ is tangent to a family of null geodesics with vanishing twist, expansion and shear ---the optical scalars--- and is a quadruple PND. The purpose of the second equation in \eqref{eq:kundt_def} is to determine the affinely parametrized character of the geodesics, since every family of non-twisting null curves is necessarily geodesic but non-affinely parametrized. The vector field $k$ is thus determined up to a scale factor constant along the null geodesics. Analogously, one can verify that the conditions \eqref{eq:kundt-N_conditions} are preserved upon conformal transformations, if and only if the conformal factor $\Omega$ is constant along each null geodesic
\begin{equation}
\pounds_k \Omega = 0.
\end{equation}

The non-twisting condition implies the existence of a family of null hypersurfaces, foliated by a family of spacelike surfaces $\Sigma$, both with normal $k$ (cf. \S 2 of \cite{kundt1961plane}). The former are called wave hypersurfaces, while the latter are called \emph{wave surfaces}. Additionally, the non-expanding and non-shearing conditions imply that the induced metric on the wave surfaces is constant along each null geodesic (cf. \S 3 of \cite{kundt1961plane}).

\subsection{Closed form and properties}

All in all, the equations in \eqref{eq:kundt_def} allow one to introduce a coordinate system $(u,v,\zeta,\bar\zeta)$ in which $k = \partial_v$ and
\begin{equation} \label{eq:kundt-N_general-metric}
ds^2 = \frac{1}{p^2} \left[ |q|^2 \! \left( h du^2 - 2 du dv + 4 \real (w du d\zeta) \right) + 2 \! \left| d \zeta \right|^2 \right],
\end{equation}
with $p,h$ real-valued functions; $q,w$ complex-valued functions; and $\partial_v p = \partial_v q = 0$ (cp. \S III of \cite{ozsvath1985plane}). The way in which \eqref{eq:kundt-N_general-metric} is written is such that the function $p$ is a conformal factor of both the spacetime metric and the induced metric on the wave surfaces, and proves useful when dealing with the type N condition, or when studying other conformal properties for that matter. The coordinate system is adapted to the specific spacetime geometry. The coordinate $u$ represents the phase of the wave and is constant on the wave hypersurfaces. On the other hand, the coordinate $v$ is an affine parameter of the null geodesics and the vectors fields $\partial_\zeta$ and $\partial_{\bar\zeta}$ are tangent to the wave surfaces. The coordinate system is determined up to transformations of the type
\begin{align} \label{eq:coord_transformations}
u = u (U), &&v = A(U) \left( V + V_0 (U, Z, \bar Z) \right), &&\zeta = \zeta (U,Z),
\end{align}
with $\dif u / \dif U > 0$, which is simply a composition of a monotonic change of the phase of the wave, an affine transformation of the null geodesics, and a holomorphic (thus conformal) transformation on the wave surfaces.

Additionally, equation \eqref{eq:type-N_def} guarantees the existence of a coordinate system such that
\begin{equation} \label{eq:function-h}
h = K v^2 + L v + H,
\end{equation}
with $K$ a constant and $L, H, w$ functions independent of $v$, all real-valued except for the complex-valued $w$, subject to
\begin{subequations} \label{eq:type-n_resulting-PDEs}
\begin{align}
&|q|^2 \, \partial_{\bar \zeta} \partial_\zeta \ln |q|^2 = -K, \label{eq:Liouvilles-eq_f}\\
&\partial_\zeta \left( L + \partial_u \ln |q|^2 - 2 i |q|^2 \, \imag (\partial_{\bar \zeta} w) \right) + 2 K w = 0 \label{eq:PDE-for-B-and-w}.
\end{align}
\end{subequations}
It is noteworthy that the first equation is a Liouville's equation, originally studied as the one satisfied by the conformal factor of two-dimensional manifolds with constant Gaussian curvature. Different representations of its general solution have been known since then \cite{liouville1853equation}.

\begin{figure}
\centering
\includegraphics[width=\columnwidth]{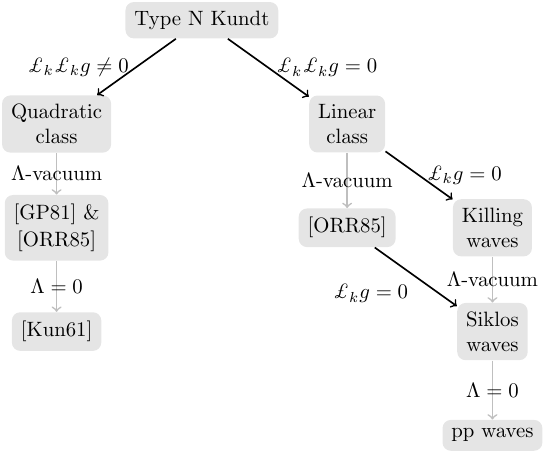}
\caption{Classification of type N Kundt spacetimes, according to the criteria \eqref{eq:quadratic-class_def}-\eqref{eq:Killing-waves_def}, and known $\Lambda$-vacuum solutions. The metric of these spacetimes can be written in the form \eqref{eq:kundt-N_general-metric} supplemented with \eqref{eq:function-h}. Closed form expressions for the metric functions of the general linear and quadratic classes are given in \eqref{eq:metric-functions_linear} and \eqref{eq:metric-functions_quadratic}, respectively, while the $\Lambda$-vacuum solutions are given in \S \ref{sec:constant-R_and_pure-radiation-S} and \S \ref{sec:profile-function_Lambda-vacuum}. The abbreviations [Kun61], [GP81], and [ORR85] stand for references \cite{kundt1961plane}, \cite{garcia1981all}, and \cite{ozsvath1985plane}, respectively.}
\label{fig:kundt-classes}
\end{figure}

Let us analyze what has been obtained so far. First, from \eqref{eq:kundt-N_general-metric}, it follows that the Gaussian curvature of the wave surfaces, calculated as half its scalar curvature, is given by
\begin{equation} \label{eq:K_wave-surfaces}
K_\Sigma = p^2 \partial_{\bar\zeta} \partial_\zeta \ln p^2.
\end{equation}
Second, all the metric components are independent of $v$ except for $g_{uu}$ which is a quadratic polynomial in $v$. Hence, one can readily see that
\begin{equation}
\pounds_k \pounds_k \pounds_k g = 0.
\end{equation}
This can happen for generic cases for which $\pounds_k \pounds_k g$ does not vanish or in special cases where it does, which in turn might happen if $\pounds_k g$ is different from zero or if it is equal to zero. This suggests the following useful classification of general type N spacetimes. We shall call the \emph{quadratic} class the one for which
\begin{align} \label{eq:quadratic-class_def}
\pounds_k \pounds_k g \neq 0, &&\partial_v^2 g_{uu} \neq 0,
\end{align}
the \emph{linear} class the one for which
\begin{align} \label{eq:linear-class_def}
\pounds_k \pounds_k g = 0, &&\partial_v^2 g_{uu} = 0,
\end{align}
and the \emph{Killing waves} class the one for which
\begin{equation} \label{eq:Killing-waves_def}
\pounds_k g = 0.
\end{equation}
As such, the quadratic and the linear classes are disjoint, while the Killing waves class is contained in the linear class (cf. figure \ref{fig:kundt-classes}).

For either the quadratic or linear classes one can find closed form general solutions to equations \eqref{eq:type-n_resulting-PDEs}. For the linear class, one can restrict the coordinate system to achieve overall
\begin{align} \label{eq:metric-functions_linear}
K = 0, &&q = 1, &&L = 2 \real (\partial _z F), &&w = -\cc{F},
\end{align}
with $F = F(u,\zeta)$. On the other hand, for the quadratic class, one can similarly restrict the coordinate system, to achieve overall
\begin{align} \label{eq:metric-functions_quadratic}
K = 1, &&q = \frac{\zeta + \bar\zeta}{\sqrt{2}}, &&L=0, &&w = \frac{\zeta + \bar\zeta}{2} \partial_\zeta^2 \left( \frac{G - \cc{G}}{\zeta + \bar\zeta} \right),
\end{align}
with $G = G(u,\zeta)$. Thus, one sees that a general type N spacetime of the Kundt class is determined by the conformal factor $p$, the function $H$, and a holomorphic function, either $F$ or $G$, all independent of $v$. The reader should note that the foregoing characterization has been made without reference to any specific matter content, neither specific field equations. This is in a similar spirit of that of \cite{podolsky2009general, coley2009kundt, ortaggio2024all}. The closed form expressions \eqref{eq:metric-functions_linear} and \eqref{eq:metric-functions_quadratic} do not seem to be reported in the literature.

For the rest of this work, it will be convenient to introduce a complex null tetrad $(k,l,m,\cc{m})$, namely one such that $g_{ab} = -2k_{(a} l_{b)}  + 2 m_{(a} \cc{m}_{b)}$, given for a general type N Kundt class spacetime by
\begin{equation} \label{eq:complex-null-tetrad}
\begin{aligned}
&k = \partial_v,\\
&l = p^2 \left\{ \frac{1}{|q|^2} \left[ \partial_u - \frac{1}{2} \left( h-2|q w|^2 \right) \partial_v \right] - 2 \real \left( w \partial_{\bar\zeta} \right) \right\},\\
&m = p \partial_{\bar\zeta}.
\end{aligned}
\end{equation}
Using this tetrad, one can readily analyze further geometric properties of the spacetime under consideration. From the Ricci identities (cf., e.g., \S 7.2 of \cite{stephani2003exact}) and \eqref{eq:kundt-N_conditions} it follows that
\begin{align} \label{eq:Rkk-and-Rkm_both-zero}
Ric (k,k) =0, &&Ric (k,m) = 0,
\end{align}
implying that $k$ is an eigenvector of the Ricci tensor $Ric$, a general feature of algebraically special spacetimes of the Kundt class. Additionally, from Gauss equation (cf., e.g., chapter 8 of \cite{lee2018introduction}) and from \eqref{eq:kundt_def} one can see that the sectional curvature of the planes spanned by $m$ and $\cc{m}$ coincides with the Gaussian curvature $K_\Sigma$ of the wave surfaces, and from the irreducible decomposition of the Riemann tensor (cf., e.g., \S 3.5 of \cite{stephani2003exact}) one can show that
\begin{equation} \label{eq:sectional-K_wave-surfaces}
K_\Sigma = S(m,\cc{m}) + \frac{R}{12},
\end{equation}
where $S$ is the traceless Ricci tensor. A similar calculation, now for the planes spanned by $k$ and $l$, again with the aid of the Ricci identities reveals
\begin{equation} \label{eq:sectional-K_wave-surfaces-perp}
S(m,\cc{m}) - \frac{R}{12} = - K \left( \frac{p}{|q|} \right)^2 + 2 |\tau|^2,
\end{equation}
where $\tau = - k_{a;b} m^a l^b$. Moreover, given that $m$ and $\cc{m}$ are tangent to the wave surfaces, these are totally geodesic if $l_{a;b} m^a m^b = 0$ and $l_{a;b} \cc{m}^a m^b = 0$ (cp. \S 4.1 of \cite{siklos1985lobatchevski}), which amounts for a general type N Kundt class spacetime to
\begin{align} \label{eq:totally-geodesic_conditions}
\partial_\zeta (|q|^2 w) = 0, &&2 \real  \left( \partial_\zeta \left[ \left( \frac{|q|}{p} \right)^2 \cc{w} \right] \right) = \partial_u p^{-2}.
\end{align}

\subsection{Spacetimes with $\Lambda$ and pure radiation matter} \label{sec:constant-R_and_pure-radiation-S}

Let us recall that a matter content of the pure radiation type (cf. \S 5.2 of \cite{stephani2003exact}) is defined by the following energy-momentum structure
\begin{equation} \label{eq:pure-radiation-matter}
T_{ab} = T(l,l) k_a k_b.
\end{equation}
Then, via Einstein's equations with a cosmological constant, spacetimes which admit that type of matter are characterized by
\begin{align} \label{eq:S-pure-radiation_R-constant}
S_{ab} = S(l, l) k_a k_b, && dR = 0.
\end{align}
For a general type N Kundt class spacetime, the foregoing represents three additional restrictions on $S$. First, one has $S(m,\cc{m}) = 0$ that together with $R$ being constant and \eqref{eq:sectional-K_wave-surfaces} imply that the Gaussian curvature of the wave surfaces is constant and given by
\begin{equation}
K_\Sigma = \frac{R}{12}.
\end{equation}
Second, $S(m,m) = 0$ together with the specific forms of $q$ serve to simplify the specific form of $p$. And third, $S(l,m) = 0$ determines the holomorphic function, $F$ for the linear class and $G$ for the quadratic class.

The linear class \eqref{eq:metric-functions_linear} only admits $R \leq 0$, which follows from \eqref{eq:sectional-K_wave-surfaces-perp}, so the wave surfaces may only have non-positive Gaussian curvature. For $R = 0$ one obtains pp waves \eqref{eq:pp-waves_metric}, while for $R < 0$ one obtains the hyperbolic-fronted waves
\begin{align} \label{eq:p-and-w_ORR}
p = \sqrt{\frac{|R|}{24}} (\zeta + \bar\zeta), &&L = i \omega (\zeta + \bar\zeta), &&w = \frac{i \omega}{2} \bar\zeta^2,
\end{align}
where $\omega$ is a constant. In the vanishing $\omega$ case they reduce to Siklos waves \cite{siklos1985lobatchevski}, while in the non-vanishing $\omega$ case, taking into account the residual coordinate freedom, one might take $\omega=1$ without loss of generality corresponding to the solutions first found in \cite{ozsvath1985plane}. On the other hand, the quadratic class \eqref{eq:metric-functions_quadratic} allows any $R \in \RR$, and hence any $K_\Sigma$, obtaining
\begin{align} \label{eq:p-and-w_GP}
p = 1 + \frac{R}{24} |\zeta|^2, &&w = 0,
\end{align}
including plane, spherical, or hyperbolic wave surfaces, found in \cite{garcia1981all, ozsvath1985plane}. Additionally, both classes feature totally geodesic wave surfaces, since they both satisfy conditions \eqref{eq:totally-geodesic_conditions}, as one can readily verify. The only undetermined function of these spacetimes, namely $H$, will be dealt with in the next section. The relations between the foregoing solutions, including how they fit into the general classes, are summarized in figure \ref{fig:kundt-classes}.

These spacetimes have the property that when writing the metric with respect to a coordinate system in which $g_{u\zeta}$ vanishes it acquires a generalized Kerr-Schild form \cite{kerr1965new,*kerr2009republication} with the background metric that of a spacetime with vanishing Weyl tensor and traceless Ricci tensor, hence locally maximally symmetric [cf. equations (4.33) and (4.36) of \cite{ozsvath1985plane}]. Moreover, conditions \eqref{eq:kundt-N_conditions} and \eqref{eq:S-pure-radiation_R-constant} imply that these spacetimes have constant curvature invariants of any order (cf. theorem 2.1 of \cite{coley2009lorentzian}), and that any symmetric tensor constructed from the metric, the curvature tensor, and its covariant derivatives is such that its trace is constant and its traceless part has a pure radiation structure, as the Ricci tensor itself, thus being \emph{almost universal} spacetimes (cf. proposition 3 and \S IV A of \cite{kuchynka2019almost}).

Remarkably, there exists conformal relations among the foregoing spacetimes of the same class (cp. \S II G and the appendix of \cite{bivcak1999gravitational}). Let $\hat g$ and $g$ be the above-mentioned spacetime metrics with vanishing and non-vanishing $R$, respectively. Then
\begin{equation} \label{eq:conformal-metric_Kundt}
\hat g_{ab} = p^2 g_{ab},
\end{equation}
is satisfied between the quadratic class spacetimes with $p$ given by \eqref{eq:p-and-w_GP} and between the Killing waves class spacetimes with $p$ given by \eqref{eq:p-and-w_ORR}. The linear class with non-vanishing $\pounds_k g$ stands out as the exception, not having a conformally related spacetime within the considered solutions.

Finally, let us show the local equivalence between the above-mentioned solutions and those found in the literature. For the linear class \eqref{eq:p-and-w_ORR} with non-vanishing scalar curvature, if one makes the transformation \eqref{eq:coord_transformations} with $\dif u/\dif U=1$, $A = 6/|R|$, and
\begin{equation}
\zeta = -\frac{e^{-i\omega U}}{\cos \left( \omega U \right)} \frac{\sqrt{\frac{|R|}{24}} e^{i\omega U} Z-1 }{ \sqrt{\frac{|R|}{24}} e^{-i\omega U} Z + 1 },
\end{equation}
allows one to choose a $V_0(U,Z,\bar Z)$ such that $g_{UZ}$ vanishes and
\begin{align} \label{eq:p-and-w_ORR_vanishing-w-gauge}
&p_{\text{new}} = 1 - \frac{|R|}{24} \left| Z \right|^2, &&q_{\text{new}} = 1+\sqrt{\frac{|R|}{24}} e^{-i\omega U} Z,
\end{align}
which, up to a re-parametrization of $\omega$ coincides with the expressions given by \cite{ozsvath1985plane} [cf. their equations (5.1) and (5.2)]. Furthermore, for the quadratic class \eqref{eq:p-and-w_GP}, if one makes simply
\begin{equation}
\zeta = e^{i \phi_0} \frac{Z+A_0 e^{i\psi_0} }{ 1-\frac{R}{24} A_0 e^{-i\psi_0} Z },
\end{equation}
the form of $p$ is preserved while
\begin{equation}
\begin{aligned}
&q_{\text{new}} = a_0 \left( 1-\frac{R}{24}  \left|Z\right|^2 \right) + 2\real \left( b_0 Z \right),\\
&\left|b_0\right|^2 +\frac{R}{24} a_0^2 = \frac{K}{2},
\end{aligned}
\end{equation}
where the constant $K$ [cf. \eqref{eq:function-h} and \eqref{eq:type-n_resulting-PDEs}] was re-inserted for the sake of comparison. The obtained expressions coincide with those of \cite{ozsvath1985plane} [cf. their equations (4.23) and (4.30)]. Transformations similar to those considered above can be found in \S IV and the appendix of \cite{ozsvath1985plane}, in \S II of \cite{bivcak1999gravitational}, and in \S 8.5 of \cite{stephani2003exact}. One could also inquire about the more compelling global equivalence of the solutions, but its analysis is more subtle and goes beyond the scope of this work.

\subsection{Wave profile function} \label{sec:profile-function_Lambda-vacuum}

For any of the spacetimes considered in the preceding section, all the non-constant and non-vanishing curvature components are determined by the function [cp. \eqref{eq:complex-null-tetrad}]
\begin{equation} \label{eq:Hprime-def}
H' = H - 2 |q w|^2,
\end{equation}
which coincides with $H$ for a vanishing $w$. The only non-vanishing Weyl scalar, $\Psi_4 = C_{abcd} \cc{m}^a l^b \cc{m}^c l^d$, is given by
\begin{equation}
\Psi_4 = -\frac{1}{2} \frac{p^4}{q^3} \partial_\zeta^2 \left( \frac{H'}{q} \right),
\end{equation}
while the only non-vanishing component of the traceless Ricci tensor is
\begin{equation} \label{eq:Sll}
S(l,l) = -\frac{1}{2} \left( \frac{p}{q} \right)^2 \Box H',
\end{equation}
where $\Box H' = {H'}_{;a}^{\ \ ;a}$. Then, for a $\Lambda$-vacuum spacetime, $H'$ satisfies the wave equation \eqref{eq:wave-eq_Hp} which, given that $\partial_v H' = 0$, yields
\begin{equation} \label{eq:eq-for-H_Lambda-vacuum}
\real \left( \partial_\zeta \left[ \left( \frac{q}{p} \right)^2 \partial_{\bar \zeta}  H' \right] \right) = 0.
\end{equation}
The general solution of the preceding equation has a closed form given by
\begin{equation} \label{eq:solution-for-H_Lambda-vacuum}
H' = \frac{2 p^3}{q} \real \left( \partial_\zeta \left( p^{-2} F \right) \right),
\end{equation}
with $F = F(u,\zeta)$ an arbitrary integration function \cite{garcia1981all,ozsvath1985plane}.

\section{Generalized conformal scalar field} \label{sec:generalized-scalar-field}

The scalar field that we will consider is coupled to gravity by the following action
\begin{widetext}
\begin{equation} \label{eq:Fernandes-action}
S[g,\Phi] = \int \dif^4 x \sqrt{|g|} \left\{ \frac{R-2\Lambda}{2\kappa} 
- \left[ \frac{R \Phi^2}{12} + \frac{ \left( \nabla\Phi \right)^2 }{2} + \lambda \Phi^4 \right]  - \frac{\alpha}{2} \left[ \GB \ln\Phi - \frac{4 G_{ab} \Phi^{;a} \Phi^{;b}}{\Phi^2} - \frac{4 \Box\Phi \left( \nabla\Phi \right)^2}{\Phi^3} + \frac{2 \left( \nabla\Phi \right)^4}{\Phi^4}  \right] \right\},
\end{equation}
\end{widetext}
with $\GB = R^2 - 4 R_{ab} R^{ab} + R_{abcd} R^{abcd}$ the Gauss-Bonnet scalar, $\left( \nabla \Phi \right)^2 = \Phi_{;a} \Phi^{;a}$ and $\Box\Phi = \Phi_{;a}^{\ \ ;a}$, $\kappa = 8\pi G$ and $\lambda, \alpha$ coupling constants. The terms enclosed in the first squared brackets correspond to the previously known conformally invariant real scalar field \cite{gursey1963reformulation,penrose1964conformal} including its self-interaction term (cf. \S 6.2 of \cite{callan1970new}), while the terms multiplied by $\alpha$ are the generalization given by Fernandes \cite{fernandes2021gravity}. The action \eqref{eq:Fernandes-action} coincides with the one given by him making $\beta_\text{there} = \kappa/6$, $\lambda_\text{there} = \kappa \lambda$, and $\alpha_\text{there} = \kappa \alpha$.

The equation of motion of the scalar field can be written as
\begin{equation} \label{eq:eom_scalar-field}
e \coloneqq \Box\Phi - \frac{1}{6} R \Phi - 4\lambda \Phi^3 - \frac{\alpha}{2} \tilde{\GB} \Phi^3 = 0,
\end{equation}
where $\tilde{\GB}$ is the Gauss-Bonnet scalar of the auxiliary metric 
\begin{equation} \label{eq:def_aux-metric}
\tilde{g}_{ab} = \Phi^2 g_{ab}.
\end{equation}
Equation \eqref{eq:eom_scalar-field} is conformally invariant, with a conformal weight $-1$ for $\Phi$. This can be seen if one rewrites it in terms of $\tilde{g}$, which is conformally invariant by construction, yielding modulo an additive constant, a linear combination with constant coefficients of the scalar curvature and the Gauss-Bonnet scalar of the auxiliary metric [cf. equations (7) and (9) of \cite{fernandes2021gravity}].
Despite being \eqref{eq:eom_scalar-field} conformally invariant, its corresponding matter action is not, suggesting the name \emph{non-Noetherian} conformal scalar field \cite{ayon2024non}. Let us remark that the contributions of $\tilde{\GB}$ to the equation of motion are of second order and quadratic in $\Phi$, as one would expect being $\tilde{\GB}$ a quadratic curvature scalar. Thus, regardless of its apparent simplicity, equation \eqref{eq:eom_scalar-field} is a fully non-linear equation, contrasting with the quasi-linear equation obtained in the vanishing $\alpha$ case. 

Let us write the obtained field equations as
\begin{equation} \label{eq:EFE_with-matter}
E_{ab} \coloneqq G_{ab} + \Lambda g_{ab} - \kappa T_{ab} = 0.
\end{equation}
Since the matter action is not conformally invariant, the energy-momentum tensor has a non-vanishing trace, as opposed to the standard conformal field. In fact, being the equation of motion conformally invariant, one can show [cf. equations (6) and (18) of \cite{fernandes2021gravity}] that the variation of \eqref{eq:Fernandes-action} with respect to the scalar field yields the identity
\begin{equation} \label{eq:trT-GB-and-eom}
\tr T - \frac{\alpha}{2} \GB \equiv \Phi e.
\end{equation}
The previous identity together with \eqref{eq:eom_scalar-field} implies, via the trace of \eqref{eq:EFE_with-matter}, that the spacetime is subject to the following geometric constraint
\begin{equation} \label{eq:geometric_constraint}
R - 4\Lambda + \frac{\kappa\alpha}{2} \GB = 0.
\end{equation}

The expression for the energy-momentum tensor is not particularly simple. However, as it happens with the standard conformal field, it can be written more compactly in terms of
\begin{equation} \label{eq:sigma-def}
\sigma = \frac{1}{\Phi}.
\end{equation}
Its traceless part is given by
\begin{align}
T_{ \langle ab \rangle } = &\frac{1}{3\sigma^3} \left[ 1 + \alpha \sigma^2 \left( 12 \Box\ln\sigma  + R \right) \right] \sigma_{\langle ;ab \rangle}\nonumber\\
& + \frac{1}{6\sigma^2} \left( 1+12\alpha \sigma^2 \Box\ln\sigma \right) S_{ab} \label{eq:TF_energy-momentum}\\
& - \frac{4\alpha}{\sigma} \left( \frac{1}{\sigma} \sigma_{ \langle ;a}^{\ \ ;c} \sigma_{;b \rangle c}^{\vphantom{;c}} + S^{ \vphantom{;c} }_{c \langle a} \sigma_{;b \rangle}^{\ \ ;c} +  C_{acbd} \sigma^{;cd} \right),\nonumber
\end{align}
%\begin{widetext}
%\begin{equation} \label{eq:TF_energy-momentum}
%\begin{aligned}
%T_{ \langle ab \rangle } = \frac{1}{3\sigma^3} \left[ 1 + \alpha \sigma^2 \left( 12 \Box\ln\sigma  + R \right) \right] \sigma_{\langle ;ab \rangle} + \frac{1}{6\sigma^2} \left( 1+12\alpha \sigma^2 \Box\ln\sigma \right) S_{ab} - \frac{4\alpha}{\sigma} \left( \frac{1}{\sigma} \sigma_{ \langle ;a}^{\ \ ;c} \sigma_{;b \rangle c}^{\vphantom{;c}} + S^{ \vphantom{;c} }_{c \langle a} \sigma_{;b \rangle}^{\ \ ;c} +  C_{acbd} \sigma^{;cd} \right),
%\end{aligned}
%\end{equation}
%\end{widetext}
%
where we have used brackets around pairs of indices to denote the symmetric traceless part, having
\begin{equation}
\begin{aligned}
&\sigma_{\langle ;ab \rangle} = \sigma_{;ab} - \frac{\Box\sigma}{4} g_{ab},\\
&\sigma_{\langle ;a}^{\ \ ;c} \sigma_{;b \rangle c}^{\vphantom{;c}} = \sigma_{;a}^{\ \ ;c} \sigma_{;bc}^{\vphantom{;c}} - \frac{\sigma_{;cd} \sigma^{;cd}}{4} g_{ab},\\
& S^{ \vphantom{;c} }_{c \langle a} \sigma_{;b \rangle}^{\ \ ;c} = S^{ \vphantom{;c} }_{c(a} \sigma_{;b)}^{\ \ ;c} - \frac{ S_{cd} \sigma^{;cd}}{4} g_{ab}.
\end{aligned}
\end{equation}
The full energy-momentum tensor can be obtained using \eqref{eq:eom_scalar-field}, \eqref{eq:trT-GB-and-eom} and \eqref{eq:TF_energy-momentum} on
\begin{equation} \label{eq:energy-momentum}
T_{ab} = T_{\langle ab \rangle} + \frac{\tr T}{4} g_{ab}.
\end{equation}

Since both the equation of motion and the energy-momentum tensor are of second order in the metric and in the scalar field, it belongs to the known general class of scalar-tensor theories \cite{horndeski1974second}. The specific choice of Horndeski's arbitrary functions which leads to the above-mentioned action were given in \cite{fernandes2021gravity} [cf. his equation (15)]. If one relaxes the condition of the energy-momentum tensor being of second order, the most general action giving rise to a conformally invariant equation of motion includes an additional term with an arbitrary function of the Weyl tensor of the auxiliary metric \eqref{eq:def_aux-metric}, as shown in \cite{ayon2024non}.

\section{Solutions} \label{sec:solutions}

We will consider the spacetimes described in \S \ref{sec:constant-R_and_pure-radiation-S}, namely type N spacetimes of the Kundt class with constant scalar curvature $R$ and whose traceless Ricci tensor has a pure radiation structure \eqref{eq:S-pure-radiation_R-constant}, sourced by the generalized conformal scalar field, restricting our attention to those fields that are constant along each null geodesic
\begin{equation} \label{eq:Phi-v-independent}
\partial_v \Phi = 0.
\end{equation}
This, motivated by the fact that it turns out to be sufficient for the energy-momentum tensor to satisfy conditions analogous to \eqref{eq:Rkk-and-Rkm_both-zero}, and thus has $k$ as an eigenvector.

Given the form of the traceless Ricci tensor, the traceless energy-momentum tensor is required to have an aligned pure radiation structure
\begin{equation} \label{eq:pure-radiation-matter_TF}
T_{\langle ab \rangle} = \left( T_{\langle cd \rangle} l^c l^d \right) k_a k_b.
\end{equation}
This amounts to the vanishing of the three components $T_{\langle ab \rangle} m^a m^b$, $T_{\langle ab \rangle} m^a \cc{m}^b$, and $T_{\langle ab \rangle} l^a m^b$, analogous to those considered at the beginning of \S \ref{sec:constant-R_and_pure-radiation-S}, supplemented with the additional requirement that
\begin{equation} \label{eq:Tll-independent-of-v}
\partial_v \left( T_{\langle ab \rangle} l^a l^b \right) = 0,
\end{equation}
given that $\partial_v \left[ S(l,l) \right] = 0$, which follows from \eqref{eq:Sll}. As it is, condition \eqref{eq:pure-radiation-matter_TF} is weaker than the one satisfied by pure radiation matter \eqref{eq:pure-radiation-matter}, since it allows a non-vanishing trace. The foregoing, together with the equation of motion \eqref{eq:eom_scalar-field}, will determine the scalar field. On the other hand, the scalar curvature $R$ is subject to the geometric constraint \eqref{eq:geometric_constraint}, while the wave profile function will be determined, once \eqref{eq:pure-radiation-matter_TF} and \eqref{eq:Tll-independent-of-v} are solved, by the field equation
\begin{equation} \label{eq:EFE_ll}
E(l,l) = 0.
\end{equation}

Before going ahead with the integration, let us analyze the particular form that the matter restriction takes when the standard conformal scalar field is considered, and its relation with that of the generalized conformal field. When $\alpha$ vanishes, it follows from \eqref{eq:S-pure-radiation_R-constant} and \eqref{eq:TF_energy-momentum} that the matter restriction \eqref{eq:pure-radiation-matter_TF} requires the traceless part of the Hessian of $\sigma$ to have also an aligned pure radiation structure
\begin{equation} \label{eq:pure-radiation-sigma-hessian_TF}
\sigma_{\langle ;ab \rangle} = \left( \sigma_{\langle ;cd \rangle}l^cl^d \right) k_a k_b.
\end{equation}
Remarkably, that structure is such that each of the last three terms in \eqref{eq:TF_energy-momentum} vanish, due to \eqref{eq:type-N_def} and \eqref{eq:S-pure-radiation_R-constant}, so it solves the matter restriction also for the non-vanishing $\alpha$ case. It will be shown below that \eqref{eq:pure-radiation-sigma-hessian_TF} characterizes one of two possible branches of the matter restriction.

\subsection{Geometric constraint} \label{sec:geometric-constraint}

\begin{figure}
\centering
\includegraphics[width=\columnwidth]{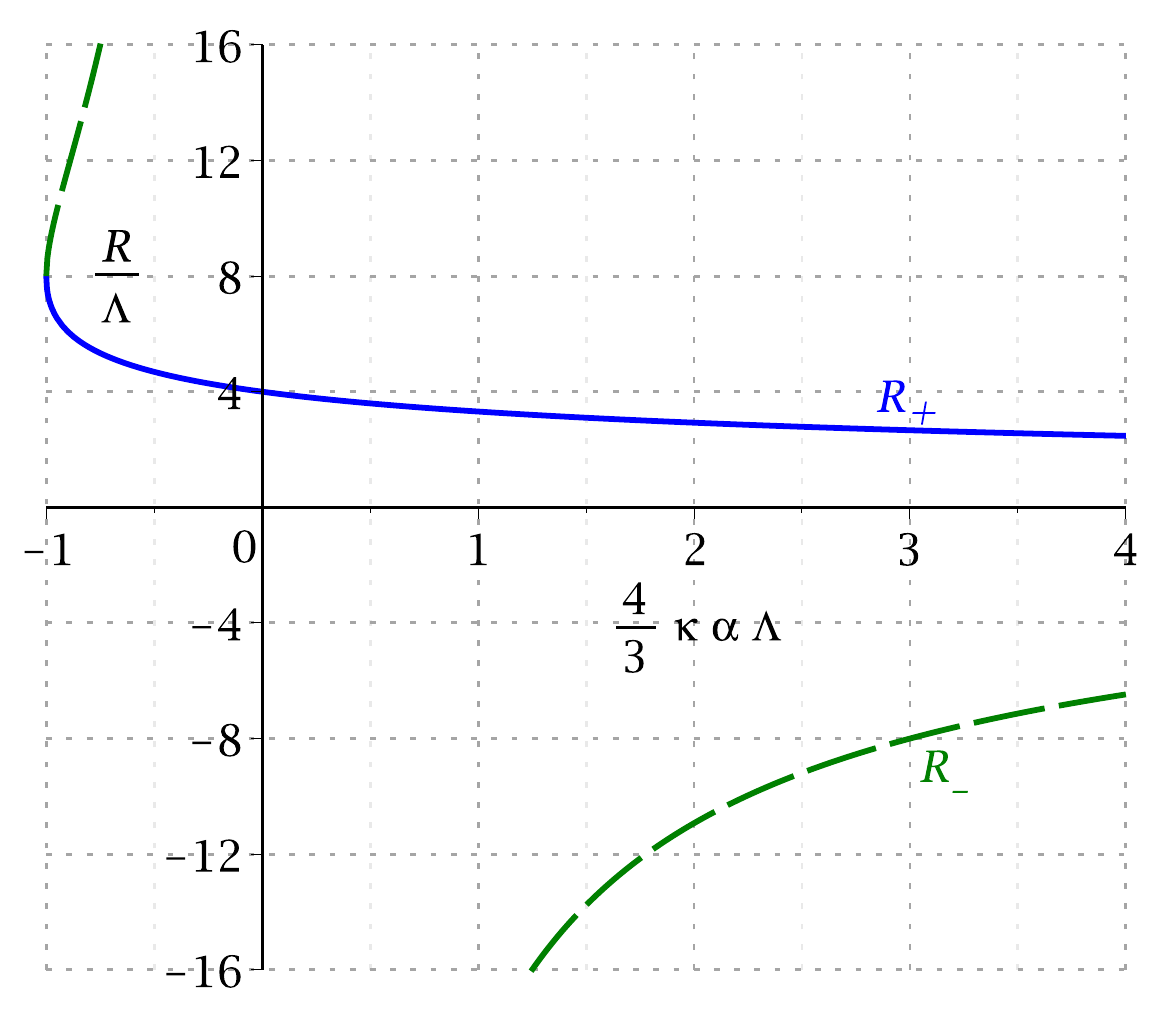}
\caption{Values for the scalar curvature as a function of $\alpha$, determined as roots of the geometric constraint \eqref{eq:geometric_constraint_type-N-Kundt} for non-vanishing $\Lambda$. The solid blue curve and the dashed green curve correspond, respectively, to the roots connected and disconnected to the standard conformal field $\alpha = 0$. For vanishing $\Lambda$, $R$ is determined by \eqref{eq:geometric_constraint_type-N-Kundt_vanishing-Lambda}.}
\label{fig:R-roots_graph}
\end{figure}

Since the type of spacetimes considered have a type N Weyl tensor and a pure radiation type traceless Ricci tensor, the Gauss-Bonnet scalar gives $\GB = R^2/6$, so the geometric constraint \eqref{eq:geometric_constraint} yields simply a quadratic equation with constant coefficients for the scalar curvature
\begin{equation} \label{eq:geometric_constraint_type-N-Kundt}
\frac{\kappa\alpha}{12} R^2 + R - 4\Lambda = 0.
\end{equation}
For generic values of the coupling constants one has two roots
\begin{equation} \label{eq:R-roots}
R = R_\pm \coloneqq \frac{6}{\kappa \alpha} \left( -1 \pm \sqrt{ 1 + \frac{4\kappa\alpha\Lambda}{3} } \right),
\end{equation}
which are real, and thus physically meaningful, only for $4\kappa\alpha\Lambda/3 > -1$. Note that only the upper sign root contains the standard scalar field as a particular case, being the only one that is finite in the limit $\alpha \to 0$, for which $R = 4\Lambda$. The two roots as a function of $\alpha$ are depicted in figure \ref{fig:R-roots_graph}. For the special case of $\Lambda = 0$ one obtains
\begin{equation} \label{eq:geometric_constraint_type-N-Kundt_vanishing-Lambda}
\left( 1 + \frac{\kappa\alpha}{12} R \right) R = 0,
\end{equation}
which in particular admits a vanishing $R$ as a solution.

For those spacetimes that are defined only for specific values of the scalar curvature, not all possible roots of \eqref{eq:geometric_constraint_type-N-Kundt} are allowed. A pp waves spacetime, which has a vanishing $R$, might only be coupled with the scalar field if $\Lambda$ vanishes. On the other hand, a linear class spacetime, which has a non-positive $R$ requires, respectively, $\Lambda < 0$ and $\alpha > 0$ for the upper and lower sign roots in \eqref{eq:R-roots}, or a vanishing $\Lambda$ and $\alpha > 0$.

\subsection{PP waves solutions}

To gain some insight, let us first consider the generalized conformal scalar field coupled with a pp waves spacetime. Substituting \eqref{eq:pp-waves_metric}, \eqref{eq:complex-null-tetrad} and \eqref{eq:Phi-v-independent} into \eqref{eq:TF_energy-momentum} yields an identity for \eqref{eq:Tll-independent-of-v} and
\begin{equation} \label{eq:TF-energy-momentum-and-N_pp}
\begin{aligned}
&T_{\langle ab \rangle} \cc{m}^a \cc{m}^b = -\frac{ N \partial_\zeta^2 \sigma }{3\sigma^3},\\
&T_{\langle ab \rangle} m^a \cc{m}^b = -\frac{ N \partial_\zeta\partial_{\bar\zeta} \sigma + 12\alpha\sigma \Hsigma }{6\sigma^3},\\
&T_{\langle ab \rangle} l^a \cc{m}^b = -\frac{ N \partial_\zeta\partial_u \sigma + 12\alpha\sigma \Husigma }{3\sigma^3},
\end{aligned}
\end{equation}
where
\begin{equation} \label{eq:N-Hsigma-and-Husigma_def}
\begin{aligned}
&N \coloneqq 24\alpha \left| \partial_\zeta\sigma \right|^2 - 1,\\
&\Hsigma \coloneqq \left| \partial_\zeta^2 \sigma \right|^2 - \left( \partial_\zeta\partial_{\bar\zeta}\sigma \right)^2,\\
&\Husigma \coloneqq \left( \partial_\zeta^2\sigma \right) \partial_{\bar\zeta}\partial_u\sigma - \left( \partial_\zeta\partial_u\sigma \right) \partial_\zeta\partial_{\bar\zeta} \sigma.
\end{aligned}
\end{equation}
Additionally, from \eqref{eq:eom_scalar-field} and \eqref{eq:trT-GB-and-eom} one obtains
\begin{equation} \label{eq:trT_pp}
\tr T = \frac{ \left( 24\alpha\sigma \partial_\zeta\partial_{\bar\zeta}\sigma - N \right) N + 96 \alpha^2\sigma^2 \Hsigma + 1 - 48\alpha\lambda }{12\alpha\sigma^4}.
\end{equation}
Furthermore, one can show that
\begin{equation} \label{eq:Hsigma-and-Husigma_N_pp}
\begin{aligned}
&\Hsigma = \frac{ \left( \partial_{\bar\zeta}\ln\partial_{\bar\zeta}\sigma \right) \partial_\zeta N - \left( \partial_\zeta\ln\partial_{\bar\zeta}\sigma \right) \partial_{\bar\zeta} N }{24\alpha},\\
&\Husigma = \frac{ \left( \partial_u\ln\partial_{\bar\zeta}\sigma \right) \partial_\zeta N - \left( \partial_\zeta\ln\partial_{\bar\zeta}\sigma \right) \partial_u N }{24\alpha}.
\end{aligned}
\end{equation}
Thus, the quantities $\Hsigma$ and $\Husigma$ are second order expressions subsidiary to $N$, in the sense that they vanish if $N$ does.

Consequently, the required condition of the traceless energy-momentum tensor having an aligned pure radiation structure \eqref{eq:pure-radiation-matter_TF} and the equation of motion of the scalar field give rise to two possible branches. One branch is determined by
\begin{align} \label{eq:N-equals-zero_and_alpha-lambda-condition}
&N = 0, &&\alpha = \frac{1}{48\lambda},
\end{align}
which are sufficient conditions for the right-hand side of \eqref{eq:TF-energy-momentum-and-N_pp} and \eqref{eq:trT_pp} to vanish, due to \eqref{eq:Hsigma-and-Husigma_N_pp}. In particular, the condition on $\alpha$ requires it to be non-vanishing, so we shall call it the \emph{strictly non-Noetherian} branch. The other branch may exist for vanishing $\alpha$, so we shall refer to it as the \emph{Noetherian-like} branch; it is characterized by the traceless part of the Hessian of $\sigma$ having a pure radiation structure [cf. \eqref{eq:pure-radiation-sigma-hessian_TF} and related discussion]. The appearance of these two branches was also identified in \cite{ayon2024cheshire} where the generalized conformal scalar field was also considered but in a different context.

\subsubsection{Noetherian-like branch}

The matter restriction \eqref{eq:pure-radiation-matter_TF} requires the components in \eqref{eq:TF-energy-momentum-and-N_pp} to vanish, or equivalently requires \eqref{eq:pure-radiation-sigma-hessian_TF}, if one entertains the possibility of a vanishing $\alpha$. This is solved by
\begin{equation} \label{eq:complete-solution_linear-sigma}
\sigma = \sqrt{\kappa} \left[ 2\real \left( b_0 \zeta \right) + c(u) \right],
\end{equation}
with $b_0$ a complex integration constant and $c(u)$ a real-valued integration function [cp. equation (12) of \cite{ayon2007higher}]. Furthermore, taking into account the residual coordinate freedom (cf., e.g., \S 2-5.4 of \cite{ehlers1962exact}) one can take $b_0$ to be real-valued and $c = 0$ without loss of generality, having
\begin{equation} \label{eq:noetherian-sol_pp-waves}
\Phi = \frac{1}{\sqrt{\kappa}} \frac{1}{ b_0 \left( \zeta + \bar\zeta \right) },
\end{equation}
where \eqref{eq:sigma-def} was used.

\begin{figure}
\centering
\includegraphics[width=\columnwidth]{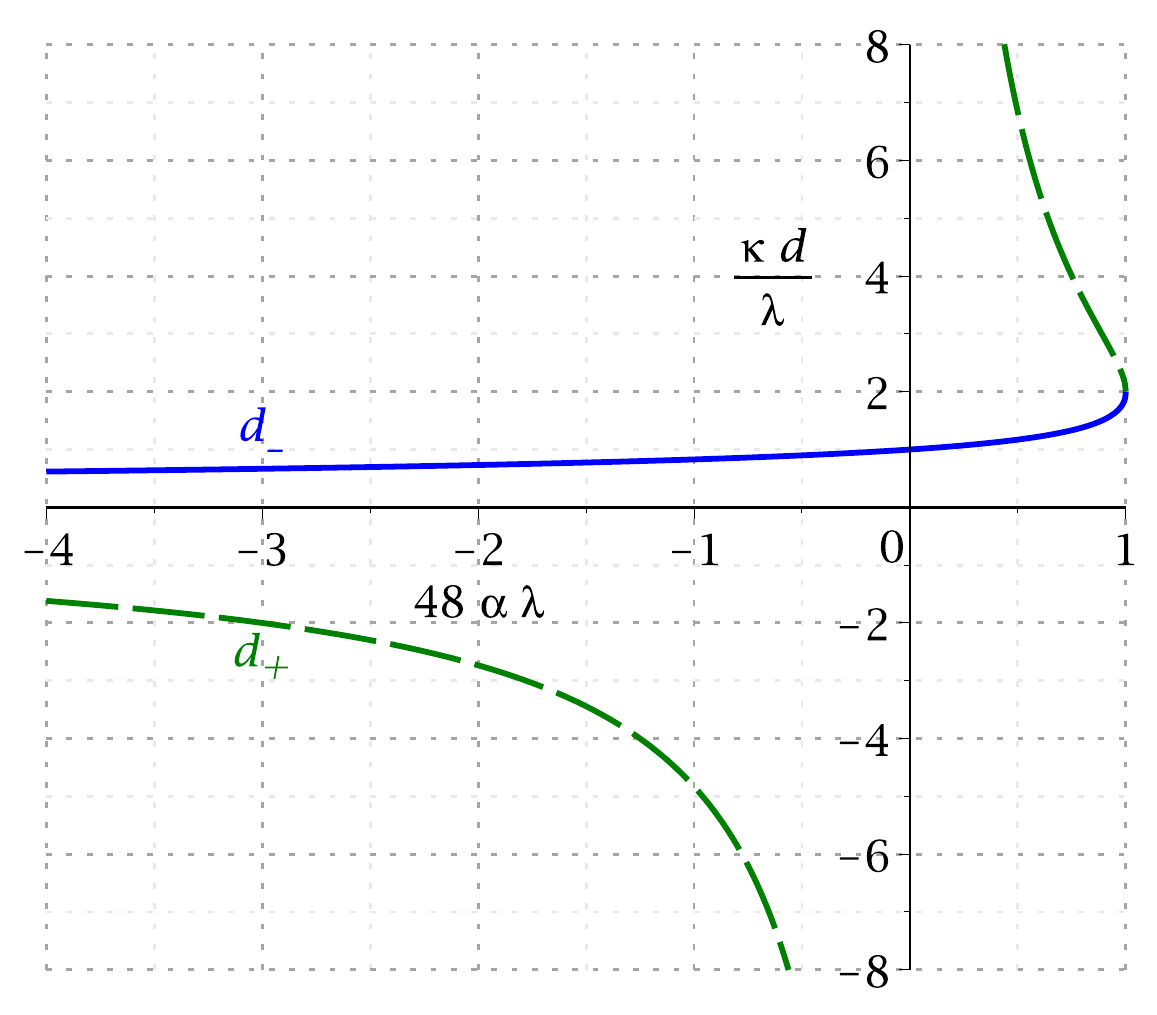}
\caption{Values for the constant $d$ restricting the integration constants of the scalar field as a function of $\alpha$, determined as roots of \eqref{eq:quadratic-eq_discrim} for non-vanishing $\lambda$. The solid blue curve and the dashed green curve correspond, respectively, to the roots connected and disconnected to the standard conformal field $\alpha = 0$. 
For vanishing $\lambda$, $d$ is determined by \eqref{eq:quadratic-eq_discrim_vanishing-lambda}.
}
\label{fig:graph_d-roots}
\end{figure}

The equation of motion requires $\tr T$ to vanish, given \eqref{eq:trT-GB-and-eom} and that $\mathcal{G} = 0$, which amounts to
\begin{equation} \label{eq:quadratic-eq_b0}
12\alpha\kappa |b_0|^4 - |b_0|^2 + \frac{\lambda}{\kappa} = 0.
\end{equation}
As with the geometric constraint, one has generically two roots
\begin{equation} \label{eq:discrim-roots-and-b0mod}
|b_0|^2 = d_\mp \coloneqq \frac{ 1 \mp \sqrt{1 - 48 \alpha \lambda} }{24 \kappa \alpha}.
\end{equation}
Certainly, the solution only exists when the right-hand side of \eqref{eq:discrim-roots-and-b0mod} is real and positive. A real-valued $d_\mp$ requires $\alpha \leq 1/(48\lambda)$, while a positive $d$ requires $\lambda > 0$ for $d_-$ and $\alpha > 0$ for $d_+$. One can readily see that only the root with the minus sign is finite in the limit $\alpha \to 0$, recovering the known solution for the standard conformal scalar field, namely $|b_0|^2 = \lambda/\kappa$ \cite{ayon2007higher}. The two roots $d_\mp$ as a function of $\alpha$ are depicted in figure \ref{fig:graph_d-roots}. For the special case of a vanishing $\lambda$ one obtains
\begin{equation}
\left( 12\alpha\kappa |b_0|^2 - 1 \right) |b_0|^2 = 0.
\end{equation}
Finally, it is worth noticing that for the critical value $\alpha = 1/(48\lambda)$ the two roots in \eqref{eq:discrim-roots-and-b0mod} coincide, $d_\mp = 2\lambda/\kappa$, allowing only positive $\lambda$.

\subsubsection{Wave profile function}

Considering the Noetherian-like branch solution for the scalar field, the field equation $E(l,l) = 0$ yields
\begin{equation} \label{eq:eq-for-H_pp-waves}
\partial_\zeta \partial_{\bar\zeta} H + \frac{\beta^2}{\zeta+\bar\zeta} \frac{\partial_\zeta + \partial_{\bar\zeta}}{\left( \zeta+\bar\zeta \right)^2 - \beta^2} = 0,
\end{equation}
with
\begin{equation}
\beta^2 = \frac{ 1-24\kappa\alpha b_0^2 }{ 6b_0^2 }.
\end{equation}

Remarkably, for the critical value $\alpha = 1/(48\lambda)$, which implies $b_0^2 = 2\lambda/\kappa$, one has $\beta = 0$ so one recovers Laplace's equation, whose general solution is given by $H = F+\cc{F}$ with $F = F(u,\zeta)$ an arbitrary integration function. The fact that one recovers the equation for the vacuum case can be understood if one analyzes the energy-momentum tensor. In fact, this critical solution yields
\begin{equation} \label{eq:pp-waves_stealth-config}
T_{ab} = 0,
\end{equation}
hence being an example of a so-called stealth configuration, first exhibited in \cite{ayon2006stealth} in the context of black holes.

For generic values of $\alpha$ equation \eqref{eq:eq-for-H_pp-waves} is not so manageable, and a closed form solution cannot be readily found. However, it does admit a separable solution when transforming to real coordinates $\zeta = x + i y$. In fact, one can readily see that the general product-separable solution is given by
\begin{equation}
H = \left( c_\mathcal{K} e^{\mathcal{K} y} + d_\mathcal{K} e^{-\mathcal{K} y} \right) \eta,
\end{equation}
with $\mathcal{K} = \mathcal{K}(u)$ a real-valued separation function and $\eta = \eta (u,x)$ satisfying
\begin{equation}
\partial_x^2 \eta + \frac{\beta^2}{2x} \frac{\partial_x \eta}{x^2-\beta^2/4} + \mathcal{K}^2 \eta = 0.
\end{equation}
The equation for $\eta$ is effectively an ordinary differential equation (ODE). It has regular singular points [cf., e.g., \S 15.3 of \cite{ince1956ordinary}] at $x = 0$ and at $x^2 = \beta^2/4$ and an irregular singular point at infinity, so it can be cast into a confluent Heun's equation \cite{slavyanov1995confluent} by making $x \mapsto (2x/\beta)^2$.

\subsubsection{Strictly non-Noetherian branch} \label{sec:pp_non-noetherian}

The branch that exists for a strictly non-vanishing $\alpha$ [cf. \eqref{eq:N-Hsigma-and-Husigma_def} and \eqref{eq:N-equals-zero_and_alpha-lambda-condition}] involves the equation $N = 0$ which is a first order non-linear partial differential equation (PDE) in the variables $\zeta$ and $\bar\zeta$. It is known that the general solution to such an equation can be constructed using as a seed a bi-parametric solution, concretely as the envelope of an arbitrary uniparametric subfamily (cf. \S I-4.2 of \cite{courant1962methods}). In fact, such a method has been used in other problems of gravitation and field theory when dealing with a first order non-linear PDE (cf., e.g., \cite{perry1997interacting,russo2024causal,ayon2024nonlinearly,ayon2024cheshire}).

Since the Noetherian-like solution is valid for any $\alpha$, it is reasonable to wonder if, when evaluated at the critical value $\alpha =1/(48\lambda)$, it solves $N = 0$. In fact, this is so, as one can readily verify by substituting \eqref{eq:complete-solution_linear-sigma} and \eqref{eq:discrim-roots-and-b0mod} into \eqref{eq:N-equals-zero_and_alpha-lambda-condition}. This comes out as a surprise, if one recalls from the discussion around \eqref{eq:N-equals-zero_and_alpha-lambda-condition} that the form of the Noetherian-like solution is determined by the system of second order quasi-linear PDEs \eqref{eq:pure-radiation-sigma-hessian_TF}, in contrast with the (single) first order fully non-linear PDE $N = 0$. Moving forward, since the latter equation does not involve $u$, it admits the solution given by \eqref{eq:complete-solution_linear-sigma} and \eqref{eq:discrim-roots-and-b0mod} even when one allows the phase of $b_0$ to be a function of $u$, namely
\begin{equation} \label{eq:separable-sol-sigma_u-dep_non-noetherian_pp-waves}
\sigma = \sqrt{2\lambda} \left[ 2\real \left( e^{i\Psi(u)} \zeta \right) + \Gamma(u) \right].
\end{equation}
Hence, this solution serves as the required bi-parametric seed solution. Following \S I-4.2 of \cite{courant1962methods}, the general solution is given by
\begin{equation} \label{eq:general-sol-sigma_non-noetherian_pp-waves}
\sigma = \sqrt{2\lambda} \left[ 2\real \left( e^{i\psi} \zeta \right) + \gamma(u,\psi) \right],
\end{equation}
with $\psi = \psi(u,\zeta,\bar\zeta)$ determined implicitly by
\begin{equation} \label{eq:eq-determining-psi_pp-waves}
\partial_\psi \gamma = -2 \imag \left( e^{i\psi} \zeta \right).
\end{equation}
The foregoing general solution is determined, for an arbitrary $\gamma$, parametrically through the function $\psi$. An explicit non-parametric representation of the general solution can be obtained by using a coordinate system which includes the function $\psi$ as a coordinate. This is more simply done making $\zeta = x + i y$, in terms of which the required coordinate transformation $(x,y) \mapsto (x,\psi)$ and the obtained solution are given by
\begin{align} \label{eq:general-sol-sigma_explicit_pp-waves}
&y = \frac{ \partial_\psi \gamma -2 x \sin\psi }{2\cos\psi}, &&\sigma = \frac{\sqrt{2\lambda}}{\cos\psi} \left[ 2x + \gamma^2 \partial_\psi \left( \frac{\sin\psi}{\gamma} \right) \right].
\end{align}
The cost one has to pay for having an explicit representation of the general solution for the scalar field is that the coordinate system is not adapted to the metric \eqref{eq:pp-waves_metric} anymore.

In summary, the strictly non-Noetherian branch has two different solutions: the first one given by \eqref{eq:separable-sol-sigma_u-dep_non-noetherian_pp-waves} which includes two arbitrary functions of $u$, and the second one given by \eqref{eq:general-sol-sigma_non-noetherian_pp-waves} and \eqref{eq:eq-determining-psi_pp-waves}, or equivalently by \eqref{eq:general-sol-sigma_explicit_pp-waves}, which exhibits an arbitrary function of two arguments $\gamma$. These two solutions contain the totality of solutions of the strictly non-Noetherian branch.

\subsection{Linear class solutions} \label{sec:linear-class-sols}

Now let us consider type N spacetimes of the Kundt class for which $\pounds_k \pounds_k g = 0$. In particular, we will consider those with hyperbolic wave fronts described in \S \ref{sec:constant-R_and_pure-radiation-S}. They are given locally by
\begin{equation}
ds^2 = \frac{1}{p^2} \left( h  du^2 - 2 du dv + 4\real \left( w \dif u \dif\zeta\right) + 2 \left| d \zeta \right|^2 \right),
\end{equation}
with $h = Lv + H$ and $L$, $p$, $w$ given by \eqref{eq:p-and-w_ORR} [cf. also \eqref{eq:p-and-w_ORR_vanishing-w-gauge} and related discussion]. We recall that without loss of generality one can take the constant $\omega$, appearing in $L$ and $w$, to be either one or zero [cf. the discussion following \eqref{eq:p-and-w_ORR}], the latter case corresponding to Siklos waves.

It proves useful to work with a conformally transformed scalar field. Taking into account the $-1$ conformal weight of $\Phi$, \eqref{eq:conformal-metric_Kundt}, and \eqref{eq:sigma-def}, one makes
\begin{align} \label{eq:sigmahat-def}
&\hat\sigma \coloneqq p \sigma, &&\Phi = \frac{p}{\hat\sigma}.
\end{align}
In terms of $\hat\sigma$ one obtains from \eqref{eq:complex-null-tetrad}, \eqref{eq:TF_energy-momentum}, and \eqref{eq:Phi-v-independent} that
\begin{equation} \label{eq:TF-T-mm-and-mmbar_N_linear}
\begin{aligned}
&T_{\langle ab \rangle} \cc{m}^a \cc{m}^b = -\frac{p^4 \hat{N} \partial_\zeta^2 \hat\sigma }{3\hat\sigma^3},\\
&T_{\langle ab \rangle} m^a \cc{m}^b = -\frac{ p^4 \left( \hat{N} \partial_\zeta\partial_{\bar\zeta} \hat\sigma + 12\alpha \hat\sigma \Hsigmahat \right) }{6\sigma^3},
\end{aligned}
\end{equation}
and
\begin{equation} \label{eq:TF-T-lmbar-and-ll-v-dependent_N_linear}
\begin{aligned}
&T_{\langle ab \rangle} l^a \cc{m}^b = -\frac{ p^5 }{12\hat\sigma^3} \left[ 4\left( \hat{N} \partial_\zeta\partial_u \hat\sigma + 12\alpha \hat\sigma \Husigmahat \right) + i \omega \varrho \right],\\
&\partial_v \left( T_{\langle ab \rangle} l^a l^b \right) = \frac{ \omega p^6 \imag \left( \partial_\zeta \hat\sigma \right) \left( 24\alpha \hat\sigma \partial_\zeta\partial_{\bar\zeta} \hat\sigma - \hat{N} \right) }{3\hat\sigma^3},
\end{aligned}
\end{equation}
where $\hat N$, $\Hsigmahat$, and $\Husigmahat$ are defined in terms of $\hat\sigma$ in the same way as their unhatted counterparts \eqref{eq:N-Hsigma-and-Husigma_def} and
\begin{align}
&\varrho \coloneqq  \left\{ 2N \left[ \zeta^3 \partial_\zeta \left( \frac{\partial_\zeta \hat\sigma}{\zeta} \right) + \bar\zeta^3 \partial_{\bar\zeta} \left( \frac{\partial_\zeta \hat\sigma}{\bar\zeta} \right) + \hat\sigma \right] \right.\\
& + \left. \hat\sigma \left[ 24\alpha\bar\zeta^2 \Hsigmahat + \left( \zeta+\bar\zeta \right) \partial_\zeta \hat{N} \right] - 24\alpha \hat\sigma^2 \partial_\zeta \left[ \partial_\zeta\hat\sigma + \partial_{\bar\zeta}\hat\sigma \right] \vphantom{\frac{\partial_\zeta \hat\sigma}{\zeta}} \right\}.\nonumber
\end{align}
In addition, from \eqref{eq:eom_scalar-field} and \eqref{eq:trT-GB-and-eom} one obtains
\begin{equation}
\begin{aligned} \label{eq:trT-GB_linear}
&\tr T - \frac{\alpha}{2} \GB = \\
&\frac{ p^4 \left[ \left( 24\alpha \hat\sigma \partial_\zeta\partial_{\bar\zeta} \hat\sigma - \hat{N} \right) \hat{N} + 96 \alpha^2 \hat\sigma^2 \Hsigmahat + 1 - 48\alpha\lambda \right] }{12\alpha \hat\sigma^4}.
\end{aligned}
\end{equation}

Modulo a global factor and the terms involving $\omega$, the preceding equations exhibit the same structure as those of pp waves. As a consequence, for vanishing $\omega$, namely for Siklos waves, the solutions for $\hat\sigma$ will be the same as those for $\sigma$ with pp waves \eqref{eq:TF-energy-momentum-and-N_pp}, with the only difference that the different residual coordinate freedom will allow other simplification of the scalar field. Conversely, for non-vanishing $\omega$ one foresees, from \eqref{eq:TF-T-lmbar-and-ll-v-dependent_N_linear}, a different behavior of the solutions; specially for the strictly non-Noetherian branch now determined by
\begin{align} \label{eq:Nhat-equals-zero_and_alpha-lambda-condition}
&\hat{N} \coloneqq 24\alpha \left| \partial_\zeta\hat\sigma \right|^2 - 1 = 0, &&\alpha = \frac{1}{48\lambda},
\end{align}
for vanishing $\omega$, while for non-vanishing $\omega$ are supplemented with the additional conditions
\begin{align} \label{eq:non-Noeth_additional-condition_non-vanishing-omega}
&\partial_\zeta^2 \sigma = 0, &&\partial_\zeta\partial_{\bar\zeta} \sigma = 0.
\end{align}

\subsubsection{Noetherian-like branch}

Let us allow $\alpha$ to vanish and analyze the solution of requiring the traceless energy-momentum tensor to be of pure radiation type, namely setting all quantities in \eqref{eq:TF-T-mm-and-mmbar_N_linear} and \eqref{eq:TF-T-lmbar-and-ll-v-dependent_N_linear} equal to zero, or equivalently requiring \eqref{eq:pure-radiation-sigma-hessian_TF}. A necessary condition is that $\hat\sigma$ has the same form as that of \eqref{eq:complete-solution_linear-sigma}. This is also sufficient for the $\omega =  0$ case while the $\omega \neq 0$ case has to be supplemented with $\cc{b_0} = b_0$ and $c = 0$. The residual coordinate freedom allows one to consider a vanishing $c$ for the $\omega = 0$ case without loss of generality [cf. equation (36) of \cite{siklos1985lobatchevski}]. Altogether, the scalar field that solves the matter restriction \eqref{eq:pure-radiation-matter_TF} is given by
\begin{equation} \label{eq:noetherian-sol_linear-class}
\Phi =
\frac{1}{\sqrt{\kappa}}
\begin{dcases}
\frac{p}{ 2\real \left( b_0 \zeta \right) } &\omega = 0,\\
\sqrt{\frac{|R|}{24}} \frac{1}{\real \left( b_0 \right)} &\omega \neq 0,
\end{dcases}
\end{equation}
where \eqref{eq:sigma-def} and \eqref{eq:sigmahat-def} were used. 
Moreover, the equation of motion requires, given \eqref{eq:trT-GB-and-eom}, that the right-hand side of \eqref{eq:trT-GB_linear} vanishes, which yields the same restriction as that for pp waves, hence restricting $|b_0|$ in terms of $\alpha$ and $\lambda$ in the exact same way [cf. \eqref{eq:quadratic-eq_b0} and related discussion].

\subsubsection{Wave profile function} \label{sec:siklos_wave-f}

Let us analyze the equation for the wave profile function obtained with the Noetherian-like branch for the scalar field. For $\omega \neq 0$, being the scalar field constant, it follows from \eqref{eq:TF_energy-momentum} that $T_{\langle ab \rangle} = \left[ 1/(6\sigma^2) \right] S_{ab}$ so the resulting field equations are the same as those for $\Lambda$-vacuum. In particular, $E(l,l) = 0$ gives \eqref{eq:eq-for-H_Lambda-vacuum}, whose general solution is \eqref{eq:solution-for-H_Lambda-vacuum}. 

Contrastingly, for $\omega=0$ the equation $E(l,l) = 0$ yields
\begin{equation} \label{eq:eq-for-H_linear}
\real \left( \partial_\zeta \left[ \frac{\partial_{\bar \zeta}  H}{\mathcal{P}^2} \right] \right) = 0,
\end{equation}
with
\begin{equation} \label{eq:p-factor_Siklos-wave-eq}
\mathcal{P} \coloneqq \frac{1}{ \sqrt{ \left( \kappa\alpha R+6 \right) - \kappa \left( 1-24\kappa\alpha d \right) \Phi^2 } } p,
\end{equation}
$d = \left|b_0\right|^2$, $p$ given by \eqref{eq:p-and-w_ORR}, $\Phi$ given by the first line of \eqref{eq:noetherian-sol_linear-class}, and recalling that $H' = H$ given that $w=0$ [cf. \eqref{eq:Hprime-def}]. The equation has schematically the same structure as the one for $\Lambda$-vacuum, with the function $\mathcal{P}$ now playing the role of $p$. Despite that, the specific coordinate dependence of $\mathcal{P}$ introduces new singular terms to the equation and represents a major obstacle when dealing with equation \eqref{eq:eq-for-H_linear}. Let us point out that the pp waves equation for $H$ \eqref{eq:eq-for-H_pp-waves} can also be written in the form given by \eqref{eq:eq-for-H_linear} and \eqref{eq:p-factor_Siklos-wave-eq}, with its corresponding $R=0$, $p=1$, and $\Phi$ given by \eqref{eq:noetherian-sol_pp-waves}.

For specific values of $\alpha$ the equation simplifies considerably and can be solved in general. In fact, for the critical value $\alpha = 1/(48\lambda)$, for which the two roots \eqref{eq:discrim-roots-and-b0mod} coincide, the function $\mathcal{P}$ ends up being a constant times $p$, so \eqref{eq:eq-for-H_linear} reduces to \eqref{eq:eq-for-H_Lambda-vacuum}, with the corresponding $q = 1$, having the same general solution \eqref{eq:solution-for-H_Lambda-vacuum}.

Similarly, for the other critical value $\alpha = -3/(4\kappa\Lambda)$, now the two roots in \eqref{eq:R-roots} coincide and the function $\mathcal{P}$ turns out to be a constant times $\hat\sigma$. Then, by making the rotation $Z = \left( b_0/\cc{b_0} \right)^{1/2}  \zeta$ one obtains $\hat\sigma = \sqrt{\kappa}  \real \left( b_0 \right) \left( Z+\bar{Z} \right)$, and the same equation as that for the linear class on $\Lambda$-vacuum is retrieved, now in terms of the rotated coordinates. Consequently, the general solution is
\begin{equation} \label{eq:solution-for-H_linear_alpha-and-Lambda-related}
H = 2 \hat\sigma^3 \real \left( \partial_Z \left( \hat\sigma^{-2} F \right) \right),
\end{equation}
where $F = F(u,Z)$ is an arbitrary integration function.

For generic values of the parameters, \eqref{eq:eq-for-H_linear} admits a separable solution when transforming to real polar coordinates $\zeta = \rho e^{i\phi}$. Indeed, the radial equation is readily solved obtaining
\begin{equation}
H = \rho \left( c_{\mathcal{K}} \rho^{\mathcal{K}} + d_{\mathcal{K}} \rho^{-\mathcal{K}} \right) \eta,
\end{equation}
with $\mathcal{K} = \mathcal{K}(u)$ a real-valued separation function and $\eta = \eta(u,\phi)$. The polar equation, namely the one satisfied by $\eta$ is formidable and does not provide any insight, so we shall not show its explicit form.

\subsubsection{Strictly non-Noetherian branch}

Considering now a strictly non-vanishing $\alpha$, the matter restriction amounts to \eqref{eq:Nhat-equals-zero_and_alpha-lambda-condition}. If $\omega = 0$, those conditions are sufficient for the right-hand side of \eqref{eq:TF-T-mm-and-mmbar_N_linear} and \eqref{eq:TF-T-lmbar-and-ll-v-dependent_N_linear} to vanish. Thus, the general solution for $\hat\sigma$ has the same form as that for $\sigma$ with pp waves (cf. section \ref{sec:pp_non-noetherian}). On the other hand, if $\omega \neq 0$, the foregoing argument does not apply, since the 
additional conditions \eqref{eq:non-Noeth_additional-condition_non-vanishing-omega} are required for the right-hand side of \eqref{eq:TF-T-lmbar-and-ll-v-dependent_N_linear} to vanish. These restrict the function $\psi$ appearing in the solution to $\partial_\zeta\psi = 0$, thus allowing only \eqref{eq:separable-sol-sigma_u-dep_non-noetherian_pp-waves} as a solution.

\subsection{Quadratic class solutions}

Finally, let us consider the type N spacetimes of the Kundt class for which $\pounds_k \pounds_k g \neq 0$ described in \S \ref{sec:constant-R_and_pure-radiation-S}. These are given locally by
\begin{equation}
ds^2 = \frac{1}{p^2} \left( q^2 \left[ (v^2 + H) \dif u^2 - 2 \dif u \dif v \right] + 2 \left| \dif \zeta \right|^2 \right),
\end{equation}
with $q$ and $p$ given by \eqref{eq:metric-functions_quadratic} and \eqref{eq:p-and-w_GP}, respectively.

As in the previous sections, it proves useful to work with the conformally transformed scalar field \eqref{eq:sigmahat-def}
\begin{align}
&\hat\sigma \coloneqq p \sigma, &&\Phi = \frac{p}{\hat\sigma}. \nonumber
\end{align}
As with the previous sections, using \eqref{eq:complex-null-tetrad}, \eqref{eq:TF_energy-momentum}, and \eqref{eq:Phi-v-independent} one obtains
\begin{equation} \label{eq:TF-T-mm_N_quadratic}
T_{\langle ab \rangle} \cc{m}^a \cc{m}^b = -\frac{p^4 \mathcal{N} \partial_\zeta^2 \hat\sigma }{3 \sqrt{2} q \hat\sigma^3},
\end{equation}
and
\begin{equation} \label{eq:TF-T-ll-v-dependent_N_quadratic}
\partial_v \left( T_{\langle ab \rangle} l^a l^b \right) = -\frac{ p^6 \left[ \frac{24}{\sqrt{2}} \alpha \hat\sigma q^3 \real \left( \partial_\zeta \left( \frac{\partial_{\bar\zeta}\hat\sigma}{q^2} \right) \right) +  \mathcal{N} \right] \partial_u\hat\sigma }{6 q^5 \hat\sigma^3},
\end{equation}
with
\begin{equation} \label{eq:N-def_quadratic}
\mathcal{N} \coloneqq 24\alpha \hat\sigma \left( \partial_\zeta\hat\sigma+\partial_{\bar\zeta}\hat\sigma \right) - \left(\zeta+\bar\zeta\right) \left( 24\alpha \left| \partial_\zeta\hat\sigma \right|^2 - 1 \right).
\end{equation}
The quantities $T_{\langle ab \rangle} m^a \cc{m}^b$, $T_{\langle ab \rangle} l^a \cc{m}^b$, and $\tr(T)-\left(\alpha/2\right) \mathcal{G}$ can be written in terms of $\mathcal{N}$ and its derivatives, analogously to those of the linear class (cf. \S \ref{sec:linear-class-sols}). The specific expressions are intricate and are not relevant for our purposes, so we shall limit ourselves only to state that
\begin{equation}
\begin{aligned}
	&\mathcal{N} = 0,\\
	&\alpha = \frac{1}{48\lambda},
\end{aligned}
\implies
\left\{
	\begin{aligned}
	&T_{\langle ab \rangle} m^a \cc{m}^b =0,\\
	&\tr(T)-\frac{\alpha}{2} \mathcal{G} = 0,\\
	&T_{\langle ab \rangle} l^a \cc{m}^b = \frac{ 4\alpha p^5 \real \left( \partial_\zeta \left( \frac{\partial_{\bar\zeta}\hat\sigma}{q^2} \right) \right) \partial_u\hat\sigma }{ \sqrt{2} q \hat\sigma \left[ \left(\zeta+\bar\zeta\right) \partial_{\bar\zeta}\hat\sigma - \hat\sigma \right] }.
	\end{aligned}
\right.
\end{equation}
Therefore, one sees the possibility of two branches, one Noetherian-like and the other strictly non-Noetherian, existing for a possibly vanishing $\alpha$ and a strictly non-vanishing $\alpha$, respectively. The latter is determined by
\begin{align} \label{eq:eqs-non-noetherian-branch_quadratic}
\mathcal{N} = 0, &&\alpha = \frac{1}{48\lambda}, &&\real \left( \partial_\zeta \left( \frac{\partial_{\bar\zeta}\hat\sigma}{q^2} \right) \right) \partial_u\hat\sigma = 0.
\end{align}

\subsubsection{Noetherian-like branch} \label{sec:noetherian-like_quadratic}

Considering the possibility that $\alpha$ vanishes, the matter restriction \eqref{eq:pure-radiation-matter_TF}, or its Noetherian-like equivalent \eqref{eq:pure-radiation-sigma-hessian_TF}, is solved by
\begin{equation} \label{eq:separable-sol-sigma_noetherian_quadratic}
\hat\sigma = \sqrt{\kappa} \left[ a_0 \left| \zeta \right|^2 + i b_0 \left( \zeta-\bar\zeta \right) + c_0 \right],
\end{equation}
with $a_0$, $b_0$ and $c_0$ real integration constants. Furthermore, the equation of motion or equivalently $\tr(T)-\left( \alpha/2 \right) \mathcal{G} = 0$, by virtue of \eqref{eq:trT-GB-and-eom}, restricts
\begin{equation} \label{eq:discrim-def}
d \coloneqq |b_0|^2 - a_0 c_0,
\end{equation}
to
\begin{equation} \label{eq:quadratic-eq_discrim}
12\alpha\kappa d^2 - d + \frac{\lambda}{\kappa} = 0.
\end{equation}
As before, one has generically two roots
\begin{equation} %\label{eq:discrim-roots}
d_\mp \coloneqq \frac{ 1 \mp \sqrt{1 - 48 \alpha \lambda} }{24 \kappa \alpha},
\end{equation}
of which only $d_-$ is finite in the limit $\alpha \to 0$, which corresponds to the standard conformal scalar field, $d = \lambda/\kappa$. Certainly, both roots require $\alpha \leq 1/(48\lambda)$ to be real and thus physically meaningful. If $\lambda$ happens to vanish, one has the special case
\begin{equation} \label{eq:quadratic-eq_discrim_vanishing-lambda}
\left( 12\alpha\kappa \, d - 1 \right) d = 0,
\end{equation}
which in particular allows a vanishing $d$ as a root.

The residual coordinate freedom allows one to set $b_0 = 0$, without loss of generality whenever $d \neq 0$, obtaining
\begin{equation} \label{eq:noetherian-Phi_quadratic-class}
\Phi = \frac{1}{\sqrt{\kappa}} \frac{p}{ c_0 - \frac{d}{c_0} \left| \zeta \right|^2 }.
\end{equation}

\subsubsection{Wave profile function}

The equation for the wave profile function, arising from $E(l,l) = 0$ with the Noetherian-like scalar field solution, is
\begin{equation} \label{eq:eq-for-H_quadratic}
\real \left( \partial_\zeta \left[ \left( \frac{q}{\mathcal{P}} \right)^2 \partial_{\bar \zeta}  H \right] \right) = 0,
\end{equation}
with $\mathcal{P}$ having the same form as that of Siklos waves \eqref{eq:p-factor_Siklos-wave-eq}
\begin{equation}
\mathcal{P} \coloneqq \frac{1}{ \sqrt{ \left( \kappa\alpha R+6 \right) - \kappa \left( 1-24\kappa\alpha d \right) \Phi^2 } } p, \nonumber
\end{equation}
but with $p$ and $\Phi$ now given by \eqref{eq:p-and-w_GP} and \eqref{eq:noetherian-Phi_quadratic-class}, respectively. 

As with Siklos waves (cf. \S \ref{sec:siklos_wave-f}), either of the critical values $\alpha = 1/(48\lambda)$ and $\alpha = -3/(4\kappa\Lambda)$ allow one to find the general solution of \eqref{eq:eq-for-H_quadratic}. For $\alpha = 1/(48\lambda)$, the function $\mathcal{P}$ ends up being a constant times $p$, so \eqref{eq:eq-for-H_quadratic} reduces to \eqref{eq:eq-for-H_Lambda-vacuum}, having the same general solution \eqref{eq:solution-for-H_Lambda-vacuum}. On the other hand, for $\alpha = -3/(4\kappa\Lambda)$, $\mathcal{P}$ turns out to be a constant times $\hat\sigma$, so \eqref{eq:eq-for-H_quadratic} acquires the same form as \eqref{eq:eq-for-H_Lambda-vacuum} with $\hat\sigma$ now playing the role of $p$, or equivalently with $R/24 \mapsto d/c_0$. Then, the corresponding general solution is
\begin{equation} \label{eq:sol-for-H_quadratic_alpha-and-Lambda-related}
H = \frac{2\hat\sigma^3}{q} \real \left( \partial_\zeta \left( \hat\sigma^{-2} F \right) \right).
\end{equation}

For generic values of $\alpha$ one has to rely on a separable solution. If one transforms to real polar coordinates $\zeta =\rho e^{i\phi}$, one has that $\partial_\phi \mathcal{P} = 0$, so \eqref{eq:eq-for-H_quadratic} allows a product-separable solution. In fact, one can see that the general product-separable solution that is single-valued is given by
\begin{equation} \label{eq:sol_product-separable_H_quadratic}
H = \frac{\real \left( \mathcal{C}_n e^{i n \phi} \right) }{\cos\phi} \eta,
\end{equation}
with $n$ a positive integer, $\mathcal{C}_n = \mathcal{C}_n(u)$ a complex-valued function, and $\eta = \eta (u,\rho)$ satisfying a second order linear ODE with respect to $\rho$, with regular singular points located at
\begin{equation}
\rho^2 \in \left\{ 0, \rho^2_-, \rho^2_+, -\frac{24}{R}, \frac{c_0^2}{d}, \infty \right\},
\end{equation}
where
\begin{widetext}
\begin{equation}
\rho^2_\pm = \frac{ 24 c_0 \left[ \left( R+144d \right) c_0 \pm \left( c_0^2 R+24d \right) \sqrt{ \left( 1-24\kappa\alpha d \right) \left( \kappa\alpha R+6 \right) }  \right] }{ 576d^2 \left( \kappa\alpha R+6 \right) - c_0^2 R^2 \left( 1-24\kappa\alpha d \right) }.
\end{equation}
\end{widetext}

When the integration constant $c_0$ happens to be
\begin{equation} \label{eq:critical-c0_quadratic}
c_0 = \sqrt{ \frac{1-24\kappa\alpha d}{ \kappa\alpha R+6 } },
\end{equation}
one has $\rho_- = 0$ so the number of (pairs of) singular points of the radial equation is reduced by one. Then, making
\begin{equation}
\xi = \left( \frac{\rho}{\rho_+} \right)^2,
\end{equation}
yields
\begin{equation} \label{eq:eq-for-H_radial-xi_quadratic}
\partial_\xi^2 \eta - \frac{2 \xi^{3/2}}{ \mathcal{P} } \left[ \partial_\xi \left( \frac{\mathcal{P}}{\xi^{3/2}} \right) \right] \partial_\xi\eta - \frac{n^2 - 1}{4\xi^2} \eta = 0,
\end{equation}
an ODE with regular singular points located at
\begin{equation}
\xi \in \left\{ 0, 1, \xi_1, \xi_2, \infty \right\},
\end{equation}
with
\begin{align}
&\xi_1 \coloneqq -\frac{ 48d \left( \kappa\alpha R+3 \right) - R }{ 2R \left( 1-24\kappa\alpha d \right) }, &&\xi_2 \coloneqq \frac{ 48d \left( \kappa\alpha R+3 \right) - R }{ 48d \left( \kappa\alpha R+6 \right) }.
\end{align}
Remarkably, for $n \neq 1$, each of its linearly independent solutions can be written as a linear combination of two hypergeometric functions with polynomial coefficients. The one that is finite at $\xi = 0$ for any $n > 0$ is given by
\begin{equation}
\begin{aligned}
\eta_1 = \xi^{\frac{r}{2}-1} & \left( p_3(\xi) F(\mathsf{a},\mathsf{b};\mathsf{c};\xi) \right.\\
& \left. \quad\quad + p_4(\xi) F(\mathsf{a}+1,\mathsf{b}+1;\mathsf{c}+1;\xi) \right),
\end{aligned}
\end{equation}
with $F(\mathsf{a},\mathsf{b};\mathsf{c};\xi)$ the hypergeometric function (cf. chapter II of \cite{erdelyi1953higher}), where
\begin{equation}
\begin{aligned}
&\mathsf{a} = \frac{ r-n+3 }{2}, &&\mathsf{b} = \mathsf{a}+n, &&\mathsf{c} = \mathsf{a} + \mathsf{b} - 2,\\
&r = \sqrt{n^2+3},
\end{aligned}
\end{equation}
and
\begin{equation}
\begin{aligned}
p_3(\xi) =& -\frac{\kappa\alpha}{3} \left[ 24\left( \kappa\alpha R+6 \right) d - R \right] \left[ \xi-1 \right]^3\\
&- \frac{\left( R-144d \right)^2}{96 Rd} \left( \xi-\frac{1}{3} \right) +2 \xi^3\\
&-\frac{\left\{ 576 \left[ r+1 \right] Rd - \left[ r+4 \right] \left[ R^2+ \left(144d\right)^2 \right] \right\} \xi^2}{ 144 Rd \left( r+2 \right) },\\
p_4(\xi) =& \frac{\left( 1-24\kappa\alpha d \right) \left( \kappa\alpha R+6 \right)}{r+1} \xi \left( \xi-1 \right) \left( \xi-\xi_1\right) \left( \xi-\xi_2 \right).
\end{aligned}
\end{equation}
The cumbersome way in which $p_3(\xi)$ is written is such that the contributions from the non-Noetherian coupling constant $\alpha$ are separated from the rest. The second solution $\eta_2$ is given by the same expression as that of $\eta_1$ but replacing $r$ for $-r$ everywhere, resulting in a divergent behavior at $\xi = 0$ for any $n$. For $n = 1$, equation \eqref{eq:eq-for-H_radial-xi_quadratic} becomes a first order ODE for $\partial_\xi \eta$ that can be solved by a straightforward integration. 

The specific value \eqref{eq:critical-c0_quadratic} of the integration constant was chosen because it simplifies considerably the radial equation and allows the hypergeometric solution. It is partially justified by considering the resulting scalar field as a \emph{simplified} representative of a homothetic-diffeomorphic class of configurations, for which both the metric and the scalar field have the same form, but not so the equation for the wave profile function.

\subsubsection{Strictly non-Noetherian branch}

Now consider the branch that exists only for non-vanishing $\alpha$. First, let us analyze the first two equations of \eqref{eq:eqs-non-noetherian-branch_quadratic}. As in the previous sections, when $\alpha=1/(48\lambda)$, the Noetherian-like solution given by \eqref{eq:separable-sol-sigma_noetherian_quadratic}, \eqref{eq:discrim-def}, and \eqref{eq:quadratic-eq_discrim} is also a solution of $\mathcal{N} = 0$, even when allowing the integration constants to be functions of $u$, namely
\begin{equation} \label{eq:separable-sol-sigma_non-noetherian_quadratic}
\hat\sigma = \frac{\sqrt{\kappa}}{ A} \left[ A^2 \left| \zeta \right|^2 + i A B \left( \zeta-\bar\zeta \right) + B^2 - \frac{2\lambda}{\kappa} \right],
\end{equation}
with $A = A(u)$ and $B = B(u)$. In fact, for this solution the quantity 
$\partial_\zeta \left( q^{-2} \partial_{\bar\zeta}\hat\sigma \right)$ is imaginary, so the third equation of \eqref{eq:eqs-non-noetherian-branch_quadratic} is also satisfied and, hence, \eqref{eq:separable-sol-sigma_non-noetherian_quadratic} is a solution of the strictly non-Noetherian branch.

Furthermore, being \eqref{eq:separable-sol-sigma_non-noetherian_quadratic} a bi-parametric solution, it serves as a seed to obtain the general solution (cf. the discussion at the beginning of \S \ref{sec:pp_non-noetherian}). Following again \S I-4.2 of \cite{courant1962methods}, the general solution is given by
\begin{equation} \label{eq:general-sol-sigma_non-noetherian_quadratic}
\hat\sigma = \frac{\sqrt{\kappa}}{a} \left[ a^2 \left| \zeta \right|^2 + i a f(a) \left( \zeta-\bar\zeta \right) + f^2(a) - \frac{2\lambda}{\kappa} \right],
\end{equation}
with $a = a(\zeta,\bar\zeta)$ a function determined by
\begin{equation} \label{eq:eq-determining-a_quadratic}
f'(a) = \frac{1}{a} \frac{ f^2(a) - a^2 \left| \zeta \right|^2 - 2\lambda/\kappa }{ 2 f(a) + i a \left( \zeta-\bar\zeta \right)}.
\end{equation}
The $u$-dependence of the functions $a$ and $f$ was excluded because the third equation of \eqref{eq:eqs-non-noetherian-branch_quadratic} rules it out. This parametrized representation of the general solution can be cast into an explicit non-parametric form by making a coordinate transformation, as done with the strictly non-Noetherian solutions of the previous sections. In terms of real coordinates $\zeta = x+iy$, the required transformation $(x,y) \mapsto (a,y)$ and the obtained solution are given by
\begin{equation} \label{eq:general-sol-sigma_non-noetherian_quadratic_explicit}
\begin{aligned}
&x = \frac{1}{a} \sqrt{ \left[ 2f(a)-ay \right] \left[ 2f(a) + a\left( y-4f'(a) \right) \right] - \frac{8\lambda}{\kappa} },\\
&\hat\sigma = \frac{\sqrt{\kappa}}{a} \left[ \left( 2f(a)-ay \right) \left( f(a)-af'(a) \right) - \frac{4\lambda}{\kappa} \right].
\end{aligned}
\end{equation}

Remarkably, and contrasting with the previous sections, there exists an additional solution not contained as a particular case of the above-mentioned general solution. This is the so-called \emph{singular solution},\footnote{This solution is singular in the following sense. Not only can it not be obtained from the general solution by any choice of the arbitrary function, but also, given the last two equations of \eqref{eq:singular-sol_eqs}, it does not satisfy conditions required in the solution of the initial value problem [cf. equation (5) in \S II-3.2 of \cite{courant1962methods}].} that is determined by the system (cf. \S I-4.3 of \cite{courant1962methods})
\begin{align} \label{eq:singular-sol_eqs}
&\mathcal{N} = 0, &&\partial_P \mathcal{N} = 0, &&\partial_{\bar P} \mathcal{N} = 0,
\end{align}
where $P = \partial_\zeta \hat\sigma$. From \eqref{eq:N-def_quadratic} and the foregoing equations it follows that the singular solution only exists for negative $\alpha=1/(48\lambda)$ and that is given by
\begin{equation} \label{eq:singular-sol-sigma_non-noetherian_quadratic}
\hat\sigma = \pm \sqrt{-2\lambda} \left( \zeta+\bar\zeta \right).
\end{equation}

In summary, the strictly non-Noetherian branch equations \eqref{eq:eqs-non-noetherian-branch_quadratic} have three different solutions: the first one given by \eqref{eq:separable-sol-sigma_non-noetherian_quadratic}, which includes two arbitrary functions of $u$; the second one given by \eqref{eq:general-sol-sigma_non-noetherian_quadratic} and \eqref{eq:eq-determining-a_quadratic}, or equivalently by \eqref{eq:general-sol-sigma_non-noetherian_quadratic_explicit}, which exhibits an arbitrary function of one argument; and the third one given by \eqref{eq:singular-sol-sigma_non-noetherian_quadratic}. These three solutions contain the totality of solutions of the strictly non-Noetherian branch.

\section{Discussion} \label{sec:discussion}

Let us begin this section by recapping the results obtained. We considered type N Kundt spacetimes with a traceless Ricci tensor having a pure radiation structure and with a constant scalar curvature [cf. \eqref{eq:S-pure-radiation_R-constant}], coupled with the generalized conformal real scalar field, which we considered to be constant along each null geodesic \eqref{eq:Phi-v-independent}. The geometric constraint \eqref{eq:geometric_constraint}, the matter restriction equations \eqref{eq:pure-radiation-matter_TF}, and the equation of motion \eqref{eq:eom_scalar-field} were solved with full generality, determining thus the value of scalar curvature and the scalar field. Two roots arose for the scalar curvature, out of which only one is finite when the coupling constant $\alpha$ vanishes, and which degenerate for the critical value $\alpha=1/(48\lambda)$. The scalar field solutions split into two roots, one Noetherian-like and the other strictly non-Noetherian, defined for possibly vanishing and for strictly non-vanishing $\alpha$, respectively. 
The Noetherian-like branch is characterized by the traceless part of the Hessian of the inverse of the scalar field having a pure radiation structure \eqref{eq:pure-radiation-sigma-hessian_TF}; this resulted in a scalar field depending only on $\zeta$ and $\bar\zeta$ for all the classes of spacetimes except for the linear class spacetime with non-vanishing $\pounds_k g$, which required a constant scalar field. The strictly non-Noetherian branch stood out being determined by a fully non-linear first order PDE. The totality of solutions of this branch includes a solution containing two arbitrary functions of the phase of the wave $u$ for all the classes [cf. \eqref{eq:separable-sol-sigma_u-dep_non-noetherian_pp-waves} and \eqref{eq:separable-sol-sigma_non-noetherian_quadratic}], a solution containing an arbitrary function $\gamma$ of two arguments for the Killing waves class \eqref{eq:general-sol-sigma_non-noetherian_pp-waves}, an arbitrary function $f$ of one argument for the quadratic class \eqref{eq:general-sol-sigma_non-noetherian_quadratic}, and a solution not containing any arbitrary element for the quadratic class \eqref{eq:singular-sol-sigma_non-noetherian_quadratic}. The resulting equation for the wave profile function with the Noetherian-like branch was shown to have a similar form as the one for $\Lambda$-vacuum [cf. \eqref{eq:eq-for-H_quadratic} and \eqref{eq:p-factor_Siklos-wave-eq} and related discussion], but including new singular terms due to the contribution of the scalar field; its general solution was found for the two critical values $\alpha=1/(48\lambda)$ and $\alpha=-3/(4\kappa\Lambda)$ while product-separable solutions were found otherwise, where different ODEs arose. In particular, the $x$ equation of a pp waves spacetime is a confluent Heun's equation, while the radial equation for the quadratic class spacetime can be cast into an equation with six regular singular points, out of which two coincide when the integration constant of the scalar field acquires a specific value, admitting as linearly independent solutions a linear combination of two hypergeometric functions with polynomial coefficients.

The generalized scalar field was shown in \S \ref{sec:geometric-constraint} to modify the value of the constant scalar curvature with respect to the $\Lambda$-vacuum case. For the root connected to the vanishing $\alpha$ case, one has $\sgn\left( R \right) = \sgn\left( \Lambda \right)$, while for the other root opposite signs of $R$ and $\Lambda$ can occur. Furthermore, for vanishing $\Lambda$, there exists the possibility of a non-vanishing $R$. Thus, the scalar field enhances, reduces, or even mimics the contribution of $\Lambda$. In particular this allows the existence of Killing waves spacetimes with positive $\Lambda$, which to our knowledge had not been exhibited so far in the literature. Of course, that only makes sense when the coupling constants yield $R<0$, required for those type of spacetimes. 
This mimicking feature is present only for a non-vanishing non-Noetherian coupling constant $\alpha$; it is, in fact, a generic feature of the theory when the non-Noetherian coupling constant $\alpha$ is non-vanishing [cf. \eqref{eq:geometric_constraint}], which in particular for the considered spacetimes is determined by \eqref{eq:geometric_constraint_type-N-Kundt}. Similar behavior has been found with $f(R)$ theories and others including couplings to $\GB$ (cf. \S V of \cite{svarc2020kundt} and \S 2 and \S 3 of \cite{baykal2022kundt}). Furthermore, an analogous behavior was shown to occur for the constant $d$ restricting the integration constants of the Noetherian-like scalar field solutions (cf., e.g., \S \ref{sec:noetherian-like_quadratic}): the contribution from the self-interaction constant $\lambda$ to $d$ is modified in the same way by the presence of a non-vanishing $\alpha$.

Another aspect exhibited of the generalized scalar field was the emergence of the strictly non-Noetherian branch. The first full characterization of this type of branch was done in \cite{ayon2024cheshire}, in a context different from that of gravitational waves. Since the ansatz we considered for the scalar field was slightly more general than that of the mentioned reference, the set of solutions allowed by the strictly non-Noetherian branch obtained here is larger. Surprisingly, as also happened in \cite{ayon2024cheshire}, the set of solutions of the two branches intersect when $\alpha = 1/(48\lambda)$. This is remarkable, considering that the  Noetherian-like branch was determined by second order quasi-linear equations, while the strictly non-Noetherian branch was determined by a first order non-linear PDE. We consider that the intersection of the two branches deserves further analysis.

It is worth noting that spacetimes of the linear class with non-vanishing $\pounds_k g$, were shown to admit a smaller family of solutions. The Noetherian-like branch only allowed a constant scalar field $\Phi$, while the strictly non-Noetherian branch only allowed a solution containing two integration functions of $u$ \eqref{eq:separable-sol-sigma_u-dep_non-noetherian_pp-waves}. This lower compatibility with the conformal scalar field can be understood by recalling that, for a given scalar curvature and wave profile function, the spacetime of the mentioned type is the only one contained in its conformal class within those type N Kundt class spacetimes of \S \ref{sec:constant-R_and_pure-radiation-S} [cf. the discussion following \eqref{eq:conformal-metric_Kundt}].

One interesting feature, uncommon in other types of analytical models for gravitational waves, is the form of the energy-momentum tensor of the scalar field. From \eqref{eq:energy-momentum}, \eqref{eq:trT-GB-and-eom} and \eqref{eq:pure-radiation-matter_TF} one obtains that the solutions presented generically satisfy
\begin{equation} \label{eq:resulting-energy-momentum}
T_{ab} = T(l,l) k_a k_b + \frac{\alpha R^2}{48} g_{ab}.
\end{equation}
The foregoing reduces to the pure radiation type \eqref{eq:pure-radiation-matter} when $R = 0$, but belongs to the more general class of energy-momentum tensors having a double null eigenvector $k$ (cf. \S 4.3 of \cite{hawking1973the}). It is characterized by the equality of the normal stresses on the wave surfaces and by an equation of state analogous to that associated with a cosmological constant (cf. \S 9.2.1 of \cite{martin2017classical}). Exceptions to 
\eqref{eq:resulting-energy-momentum} are the linear class spacetime with non-vanishing $\pounds_k g$ and pp waves spacetimes for a generic $\alpha$, both of which resulted in pure radiation matter \eqref{eq:pure-radiation-matter}. Another exception is the pp waves spacetime at the critical value $\alpha=1/(48\lambda)$, which turned out to be a stealth configuration \eqref{eq:pp-waves_stealth-config}. For \eqref{eq:resulting-energy-momentum}, the null energy condition requires $T(l,l)>0$ while the weak energy condition requires additionally $\alpha \leq 0$. Interestingly, the emergence of an energy-momentum tensor with a null eigenvector in scalar-tensor theories was foreseen in \cite{banerjee2023realizations}, but one with a triple, rather than a double, null eigenvector.

Finally, let us comment on the considered ansatz for the scalar field \eqref{eq:Phi-v-independent}, namely one that is constant along each null geodesic. Our main motivation was that it is a sufficient condition for the energy-momentum tensor to have $k$ as an eigenvector. Additionally, this enabled a straightforward integration of the matter restriction equations \eqref{eq:pure-radiation-matter_TF}. However, whether a type N Kundt spacetime can be coupled with a conformal scalar field with arbitrary dependence remains an open question. We believe this will end up being incompatible with the type N character requiring the spacetime to be conformally flat, but further investigations are required.

\section{Acknowledgments}

The author is grateful to E. Ayón-Beato for drawing his attention to this problem and for constructive criticism during the realization of this work. Gratitude is due also to M. Hassaine and D. Flores-Alfonso for insightful discussions and valuable suggestions, and to two anonymous referees, whose inquiries enhanced this paper. In verifying many of the calculations done, the use of \emph{Maple}'s package \mbox{\emph{DifferentialGeometry}} \cite{anderson2012new} proved quite useful. The author was supported by Secihti, former Conahcyt, fellowship ``Estancias Posdoctorales por M\'exico'' agreement I1200/320/2022. Partial support from Secihti grant A1-S-11548 is also acknowledged.

\bibliography{refs}

\end{document}